\documentclass[12pt,preprint]{aastex}

\shorttitle{Effective destruction of CO by cosmic rays}
\shortauthors{Bisbas et al.}

\def\mean#1{\left< #1 \right>}

\begin{document}

\title{Effective destruction of CO by cosmic rays: implications for tracing H$_2$ gas in the Universe}

\author{Thomas G. Bisbas}
\affil{Department of Physics and Astronomy, University College London, Kathleen Lonsdale Building, Gower Place, WC1E 6BT, London}
\email{t.bisbas@ucl.ac.uk}

\author{Padelis P. Papadopoulos}
\affil{School of Physics and Astronomy, Cardiff University, Queen's Buildings, The Parade, Cardiff, CF24 3AA, UK}
\affil{Research Center for Astronomy, Academy of Athens, Soranou Efessiou 4, GR-115 27 Athens, Greece}

\and

\author{Serena Viti}
\affil{Department of Physics and Astronomy, University College London, Kathleen Lonsdale Building, Gower Place, WC1E 6BT, London}

\begin{abstract}

We report on  the effects of cosmic rays (CRs) on  the abundance of CO in $\rm H_2$ clouds under conditions typical for star-forming galaxies in the  Universe. We  discover that this most important  molecule for tracing  H$_2$ gas is  very effectively destroyed in  ISM environments with CR energy densities $\rm U_{CR}\sim(50-10^{3})\times U_{CR,Gal}$, a  range expected in  numerous  star-forming  systems throughout  the Universe. This density-dependent effect operates volumetrically rather than only on molecular cloud  surfaces (i.e. unlike FUV radiation that also destroys CO), and is facilitated by: a) the direct destruction of CO  by  CRs,  and  b)  a reaction  channel activated  by  CR-produced He$^{+}$. The effect  we uncover is strong enough  to render Milky-Way type   Giant  Molecular Clouds   (GMCs)  very   CO-poor  (and   thus CO-untraceable), even  in   ISM  environments  with  rather  modestly enhanced average CR energy densities of $\rm U_{CR}\sim(10-50)\times\rm U_{CR,Gal}$. We conclude  that the CR-induced destruction  of CO in molecular clouds,  unhindered by dust absorption, is  perhaps the  single most important  factor controlling the  CO-visibility   of  molecular  gas   in vigorously  star-forming galaxies.   We  anticipate that  a second  order  effect of  this  CO destruction mechanism  will be to  make the H$_2$ distribution  in the gas-rich disks  of such galaxies appear much  clumpier in  CO $J$=1--0, 2--1 line emission than it actually is.  Finally we give an analytical approximation of  the CO/H$_2$ abundance ratio as a function of gas density and CR  energy density for use in  galaxy-size or cosmological hydrodynamical  simulations, and propose  some  key observational tests.

\end{abstract}

\keywords{(ISM:) cosmic rays -- astrochemistry -- methods: numerical -- ISM: abundances -- (ISM:) photon-dominated region (PDR) -- galaxies: ISM}

\section{Introduction}

Using the mm/submm rotational transitions of CO, the most abundant molecule after H$_2$ (CO/H$_2\sim 10^{-4}$), as tracers of the H$_2$ distribution and average physical conditions in galaxies is now a well-established method, and indeed a main driver behind the largest ground-based astronomical facility on Earth, the Atacama Large Millimeter Array (ALMA) in Llano Chajnantor of North Chile. Ever since the first detection of a bright CO $J=1-0$ line in Orion \citep{Wils70} there has been tremendous progress towards detecting ever higher-$J$ transitions of CO, in our own Galaxy as well as in others \citep[e.g.][]{Cron76, Gust93, Deve94, Dumk01, Yao03, Yao03e, Mao10, Papa07, Papa10a, Leec10} and of other molecules such as HCN, HCO$^+$, and CS \citep{Nguy92, Solo92, Pagl97, Gao04, Grac06, Grac08, Papa07S, Krip08, Grev09, Zhan14}. The relative strengths of such lines can then be used as powerful probes of the average density, temperature, and dynamical state (i.e. gravitationally bound or unbound) of the molecular gas in galaxies \citep[][ and references therein]{Papa12a}. A lot of effort has been focused particularly towards using the two lowest CO rotational lines as H$_2$ gas mass tracers in galaxies via the so-called $X_{\rm CO}$-factor \citep[e.g.][]{Dick86, Malo88, Youn91, Solo92, Brya96, Solo97, Down98}. In this context higher-$J$ CO and other molecular lines remain very important since the constrains they set on the average density, thermal and dynamical state of H$_2$ cloud ensembles\footnote{$X_{\rm CO}$ retains its mass-tracing value only for cloud ensembles rather than cloud parts or single H$_2$ clouds} help determine $X_{\rm CO}$ and find whether it deviates from its Galactic calibrated value $X_{\rm CO,Gal}\sim5\,X_{\ell}$\footnote{$X_{\ell}={\rm M_{\odot}(K\,km\,s^{-1}\,  pc^2)^{-1}}$} \citep[see][for a recent such study]{Papa12b}.

Nevertheless, all available studies quantifying the effects of the H$_2$ physical conditions on the (H$_2$ mass)/(CO luminosity) ratio rely on the premise of a CO-rich H$_2$ gas phase. Significantly lower CO/H$_2$ ratios would render the use of CO lines  as molecular gas mass tracers   problematic.  Extensive observational and theoretical work has indeed shown that a CO-poor phase becomes possible for ISM with low  average metallicities (Z$\la0.2Z_{\odot}$)  and strong FUV fields \citep[e.g.][]{Malo88, Madd97, Isra97, Pak98, Bell06a}.  Metallicity (Z)  and the average FUV radiation  field ($\rm G_{\circ}$) then emerge as the two most important factors determining CO/H$_2$ in metal-poor ISM, while models  that incorporate the  underlying ISM physics  have been  made  to  correct for such CO-dark H$_2$  mass  per  cloud \citep{Pak98, Bola99}. These can be of help as long as: a)  some CO survives in the inner cloud regions (so that CO line emission can still mark such clouds),  b) the average metallicity and mean  FUV radiation field are known (so that their values  can be inserted  in the models). Such models  will obviously not work for metallicities  that fall too  low and/or average FUV  fields that are strong  enough to keep CO  fully dissociated\footnote{For clarity we must emphasize here that even in the CO-rich parts of the Galaxy    CO-poor regions  do exist (e.g.  diffuse H$_2$ clouds, outer parts    of GMCs, regions very near O,  B stars), but their masses are typically too small    to make any impact on the  total H$_2$ mass budget as traced by the CO-rich H$_2$ gas.} \citep[e.g.][]{Bola99, Bell06b}.

 Despite existing theoretical work that investigated the effects of CRs on ISM chemistry \citep[e.g.][]{Baye11,Meij11}, the effects on  CO/H$_2$  of CRs (which, unlike FUV radiation,  travel  nearly unimpeded through the irrespective  of its dust content),  or their impact on the observability of H$_2$ gas via CO lines has not  been investigated in detail.  Recently, \citet{Bial14}  presented an analytical study of the importance of CRs in the abundance of the most important species under different metallicities,  using isothermal calculations. In this work, we present our investigation of the effects produced by CRs starting with  the ISM  of metal-rich  star-forming galaxies but, unlike \citet{Bial14}, we perform full thermochemical calculations (i.e. we solve for thermal balance) while using a much more extensive chemical network. We assume CRs fully penetrating the molecular clouds, with an average  CR energy density  scaling as $\rm U_{CR}$$\propto\rho_{SFR}$ \citep[where $\rho_{\rm SFR}$ is the average star-formation rate energy density; see][and references therein for detailed arguments]{Papa10a}.

\section{The effect of CRs on the CO/H$_2$ ratio in molecular clouds}
\label{sec:effectCR}

We first perform a grid of runs of one-dimensional uniform density clouds in which we vary the density, $n_{\rm H}$, from $(10^2-10^5)\,{\rm cm}^{-3}$ and the cosmic ray ionization rate, $\zeta_{\rm CR}$, from $10^{-17}-10^{-13}\,{\rm s}^{-1}$. The choice for the $\rm \zeta _{CR}$ range is  dictated by  that deduced within the Milky  Way   ($\sim (1-5)\times 10^{-17}$\,s$^{-1}$), and from the $\rm \zeta _{CR}$  expected    in galaxies with star-formation (SF) rate  densities of  $\rm \rho_{SFR}$$\sim(1-10^3)\times\rho _{\rm SFR,Gal}$     (assuming $\rm  \zeta _{CR}\propto U_{CR}$ where ${\rm U_{CR}}$ is the CR energy density, and      $\rm U_{CR}\propto \rho_{SFR}$ for a CR-transparent ISM). 

\begin{figure*}[htbp]
  \centering
\includegraphics[width=0.9\textwidth]{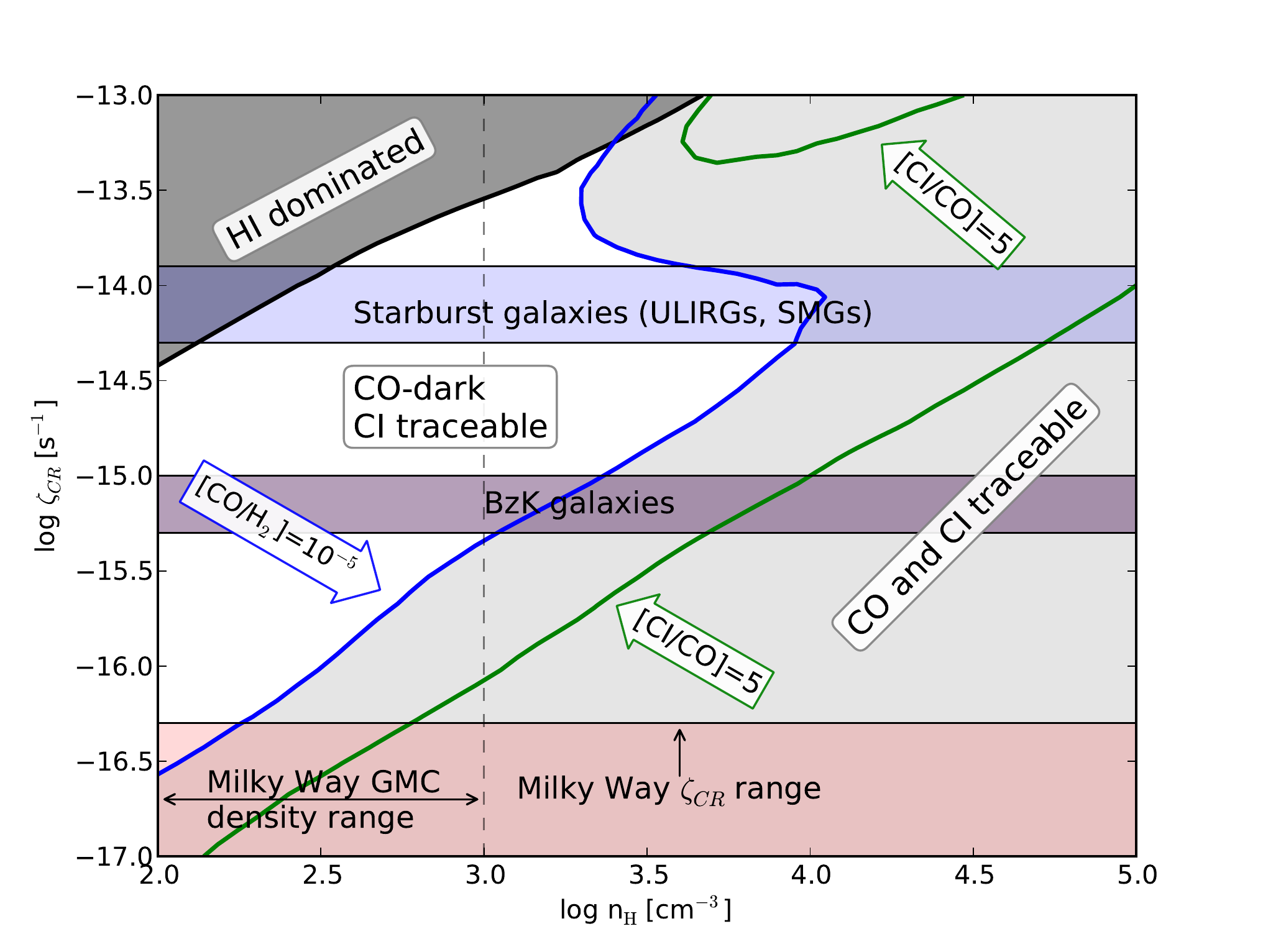}
\caption{ Map showing the traceability of H$_2$ using CO and C{\sc i} as a function of H number density $n_{\rm H}$ and the cosmic-ray ionization rate $\zeta_{\rm CR}$. The three horizontal areas correspond to the $\zeta_{\rm CR}$ range for (from bottom to top) the Milky Way, BzK galaxies, and Starburst galaxies. The shaded area below the blue line (CO/H$_2=10^{-5}$) corresponds to the region for which CO and C{\sc i} can both trace H$_2$. The region above the blue line and below the top left dark-shaded region corresponds to the area in which CO has been destroyed by cosmic-rays but H$_2$ survives, making C{\sc i} the only possible tracer. The dark-shaded region on top left corresponds to the area where H$_2$ is also destroyed by the interaction of cosmic rays hence being rich in H{\sc i} (H{\sc i}/2H$_2=1$). Note that for $\zeta_{\rm CR}>10^{-14}\,{\rm s}^{-1}$ CO is destroyed less effectively for lower densities than $n_{\rm H}=10^{4}\,{\rm cm}^{-3}$ due to the destruction of He$^+$. In \S\ref{ssec:chemistry} we discuss the chemistry behind this behaviour.}
\label{fig:zcr}
\end{figure*}

In order to isolate the  effects of CRs on the  important CO/H$_2$ ratio we set the incident FUV field $G=0G_{\circ}$ \citep[$G_{\circ}$  the average interstellar radiation field as described by][]{Drai78}, and  consider only Solar metallicity (Z=$\rm  Z_{\odot}$) gas.  We chose  a  density range  that  encompasses  the low-density  Galactic regions where the  H{\sc i}$\rightarrow$$\rm H_2$ phase  transition takes place  in   the  Cold   Neutral  Medium ($\langle n_{\rm H}\rangle  $$\sim $200\,cm$^{-3}$), to typical Galactic GMCs ($\langle n_{\rm H}\rangle $$\sim $$500-10^3$\,cm$^{-3}$),  and up to the high densities typical  of SF cores  ($\langle n_{\rm H}\rangle  $$\sim $$10^5$\,cm$^{-3}$). We  then  use  the {\sc 3d-pdr}   code (see Appendix \ref{app}) to  conduct  full  thermochemical calculations on a uniform density  cloud permeated by an average    $\rm U_{CR}$ (that proportionally sets  $\zeta _{CR}$) for the grid of $[n_{\rm H}-\zeta_{\rm CR}]$ values. The  results on  the CO/H$_2$    ratio (blue lines, marking CO/H$_2=10^{-5}$) can  be viewed in Fig. \ref{fig:zcr}, where we also  show the  C{\sc i}/CO    ratio (green  lines, indicating the C{\sc i}/CO$=\!5$ ratio) since  atomic carbon  is  the  dominant   species  resulting  from  the     CR-induced destruction of CO.

From Fig.\ref{fig:zcr}  it can be seen that  molecular clouds  with densities typical of the Galactic GMC density range ($\sim $$10^{2}-10^{3}$\,cm$^{-3}$)  can become very  CO-poor even in ISM  environments with  SFR densities  (and thus  $\rm  U_{CR}$) only $\sim10\times$Galactic.  This  is even  more the case for  the ISM  of gas-rich  SF systems  like  the so-called  BzK  galaxies.  These  are isolated  disks,  typically at  z$\sim1-2$,  with  ${\rm SFR}\sim(50-150)\times $Galactic,  and  representative  of a major SF mode in the Early Universe  \citep{Dadd10}.  In such systems CO/H$_2<10^{-5}$  (10  times  less abundant  than  in  the Galaxy) {\it for much of the  range  of average {\rm H}$_2$ gas  densities typical for  Galactic GMCs.}  Thus,  unless the molecular gas  in such galaxies    lies   preferentially    in    cloud   structures    with $n({\rm H}_2)>10^3\,{\rm cm}^{-3}$, CO imaging for the purpose of revealing H$_2$ gas disk sizes, average molecular gas physical conditions, and H$_2$ mass distribution \citep[e.g. ][]{Dadd10,Dadd14} may be inadequate.   Figure \ref{fig:zcr} also reveals that efficient CR-induced  CO destruction encompasses  ever increasing gas densities for  higher $\rm  \zeta _{CR}$, reaching  up to  $n\sim 10^4\,{\rm cm}^{-3}$ for $\rm \zeta _{CR}$ values expected in compact merger/starburst galaxies  in  the local  Universe  (ULIRGs), and similar systems  like the so-called Submm Galaxies (SMGs)  found at much larger numbers in the distant Universe \citep{Smai97,Hugh98}

Furthermore, even in the Galaxy there seems to be a $[n_{\rm H}-\zeta_{\rm CR}]$ parameter space where low-density molecular gas ($n_{\rm H}\sim (1-3)\times10^{2}\,{\rm cm}^{-3}$) can  be very CO-poor. Such a gas phase may indeed exist in the outer Galactic disk ($R_{\rm Gal}\geq15\,{\rm kpc}$)  where  a  combination of  low metallicities ($Z\sim 0.15-0.2 Z_{\odot}$) and low densities keep CO formation rates  low while  a still  substantial  FUV radiation field  maintains high  CO destruction rates.  This can yield a CO-dark H$_2$ phase that can be quite  massive \citep{Papa02}.  Any additional CO destruction due to CRs (which propagate freely  to the  outer Galactic disk) can only further reduce  CO/H$_2$, rendering the corresponding molecular gas phase even harder to detect via CO line emission \citep[see also][for a recent study of CO-dark H$_2$ gas in the Galaxy]{Wolf10}.

\subsection{The intensity of  CO $J$=1--0, 2--1 and C{\sc i} 1--0, 2--1 lines}
\label{ssec:intensity}
 
We will now examine the emergent  intensities of the CO $J$=1--0, 2--1 ($\sim $115\,GHz, $\sim $230\,GHz) and  C{\sc i} $J$=1--0, 2--1 ($\sim $492\,GHz, $\sim $809\,GHz) lines across the $\rm [n_H-\zeta _{CR}]$ parameter space shown in Fig. \ref{fig:zcr}, in order to access their detectability in CO-poor versus CO-rich gas phases. We choose only the two lowest-$J$ CO lines as the only CO transitions that are not excitation-biased, when it comes to the underlying state of the gas necessary to significantly excite them. Higher-$J$ CO lines will trace mainly warm and dense H$_2$ gas ($n\ga 10^{4}$\,cm$^{-3}$, $\rm E_J/k_B\ga 30\,K$), and are not reliable tracers of the general H$_2$ gas distribution irrespective of its state. Thus  $J$=1--0, 2--1 are the CO lines where most observational calibrations of the so-called $\rm X_{CO}$=$\rm M(H_2)/L_{CO}$ factor have been performed \citep[see][for a review]{Bola13}. The two C{\sc i} lines on the other hand are  powerful alternatives  for tracing the general H$_2$ gas distribution (especially C{\sc i} $J$=1--0), often being brighter than CO $J$=1--0, 2--1 in CO-rich gas while remaining luminous even in CO-poor gas \citep{Papa04}. 

\begin{figure}[htbp]
  \centering
\includegraphics[width=0.45\textwidth]{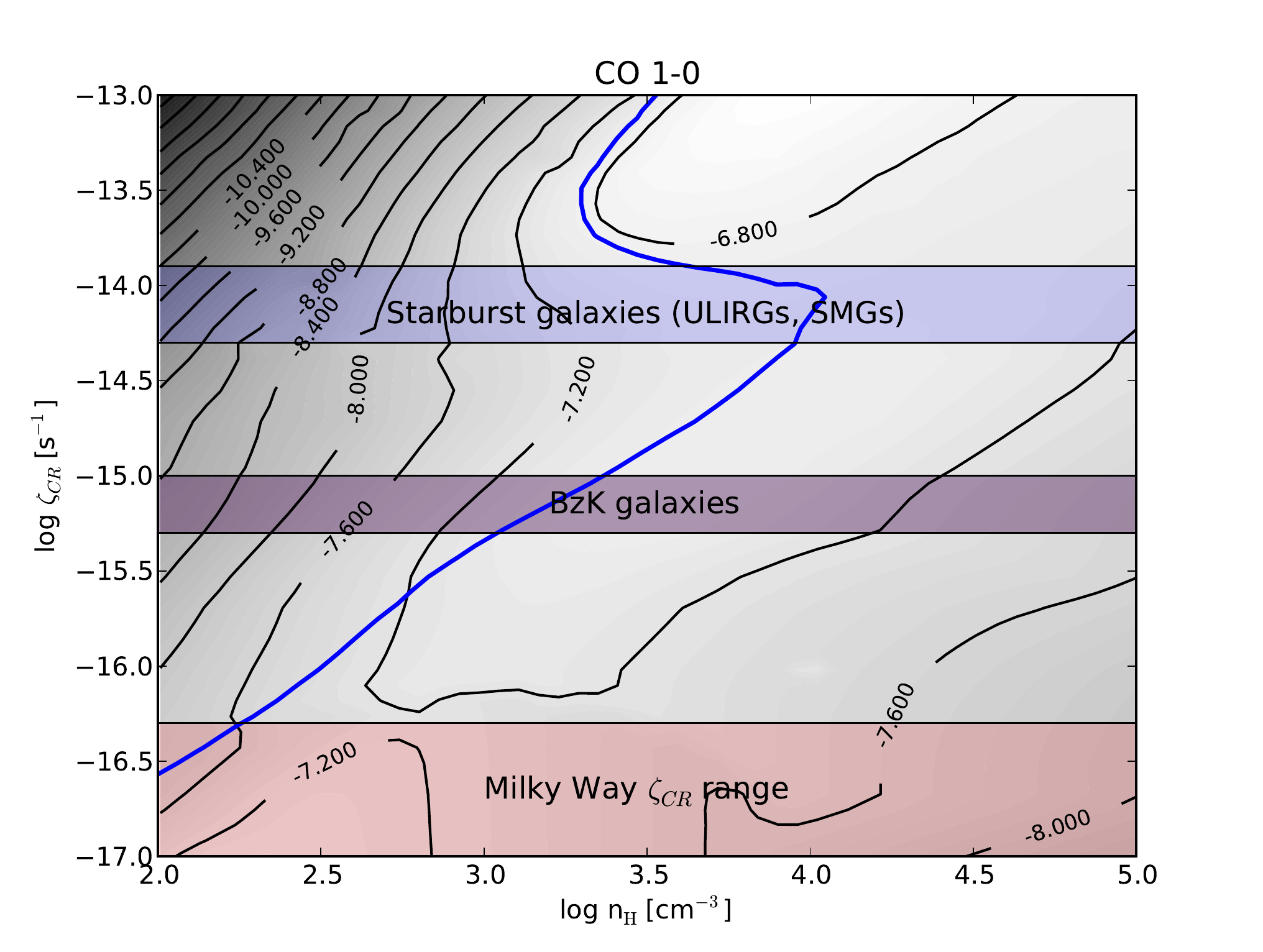}
\includegraphics[width=0.45\textwidth]{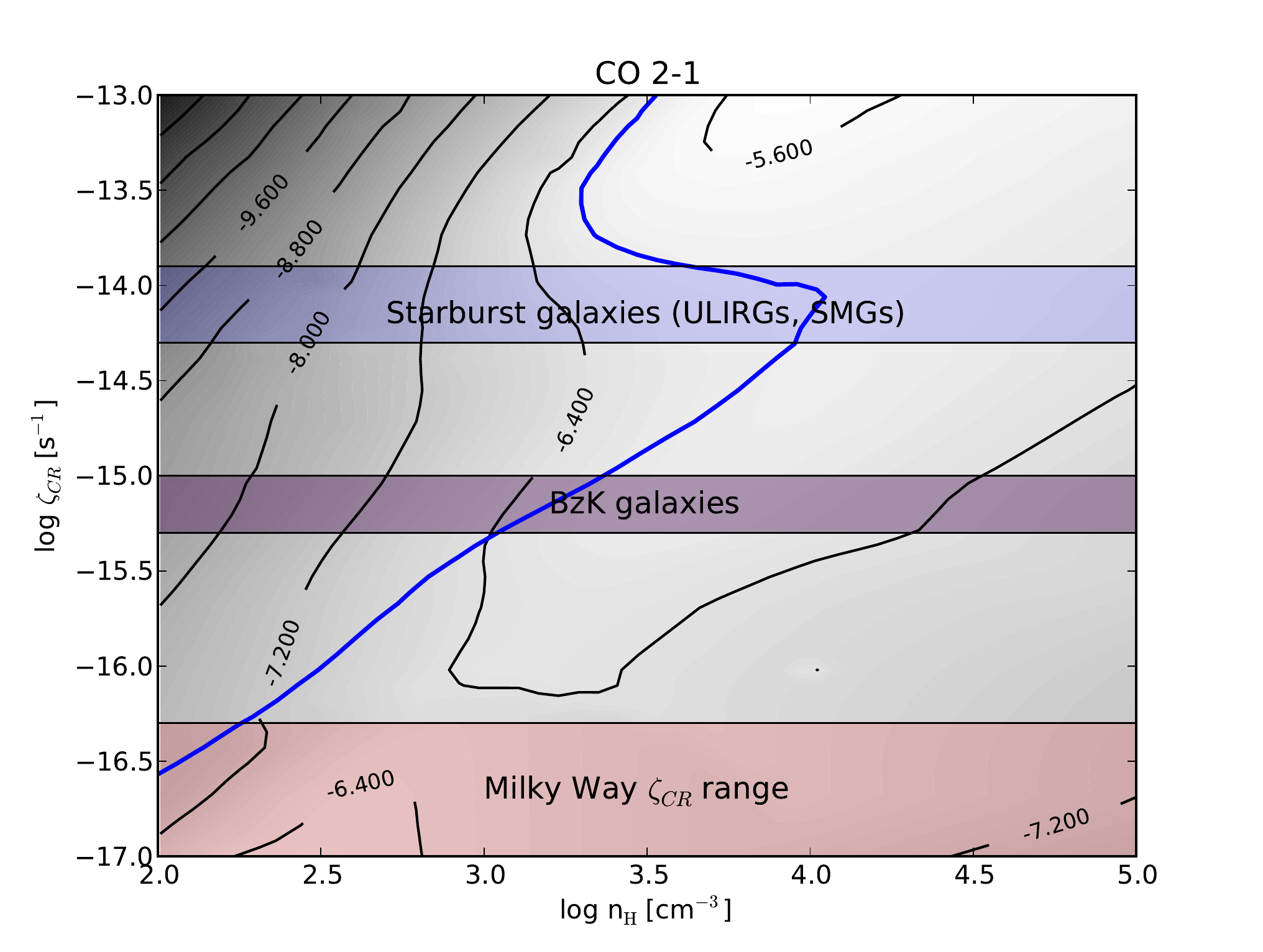}
\caption{ Logaritmic contour/grayscale plots showing the brightnesses of CO $J$=1--0 (left) and CO $J$=2--1 (right) in the $[n_{\rm H}-\zeta_{\rm CR}]$ space. The brightness is calculated by integrating the local emissivity for a depth of $A_{\rm V}=10\,{\rm mag}$, and is in units of ${\rm erg}\,{\rm cm}^{-2}\,{\rm s}^{-1}\,{\rm sr}^{-1}$. The blue line corresponds to CO/H$_2=10^{-5}$ (see Fig. \ref{fig:zcr} for further details). The large reduction of the CO line strengths in the CO-poor regions is evident (see \S\ref{ssec:intensity} for detailed discussion).}
\label{fig:CObright}
\end{figure}

\begin{figure}[htbp]
  \centering
\includegraphics[width=0.45\textwidth]{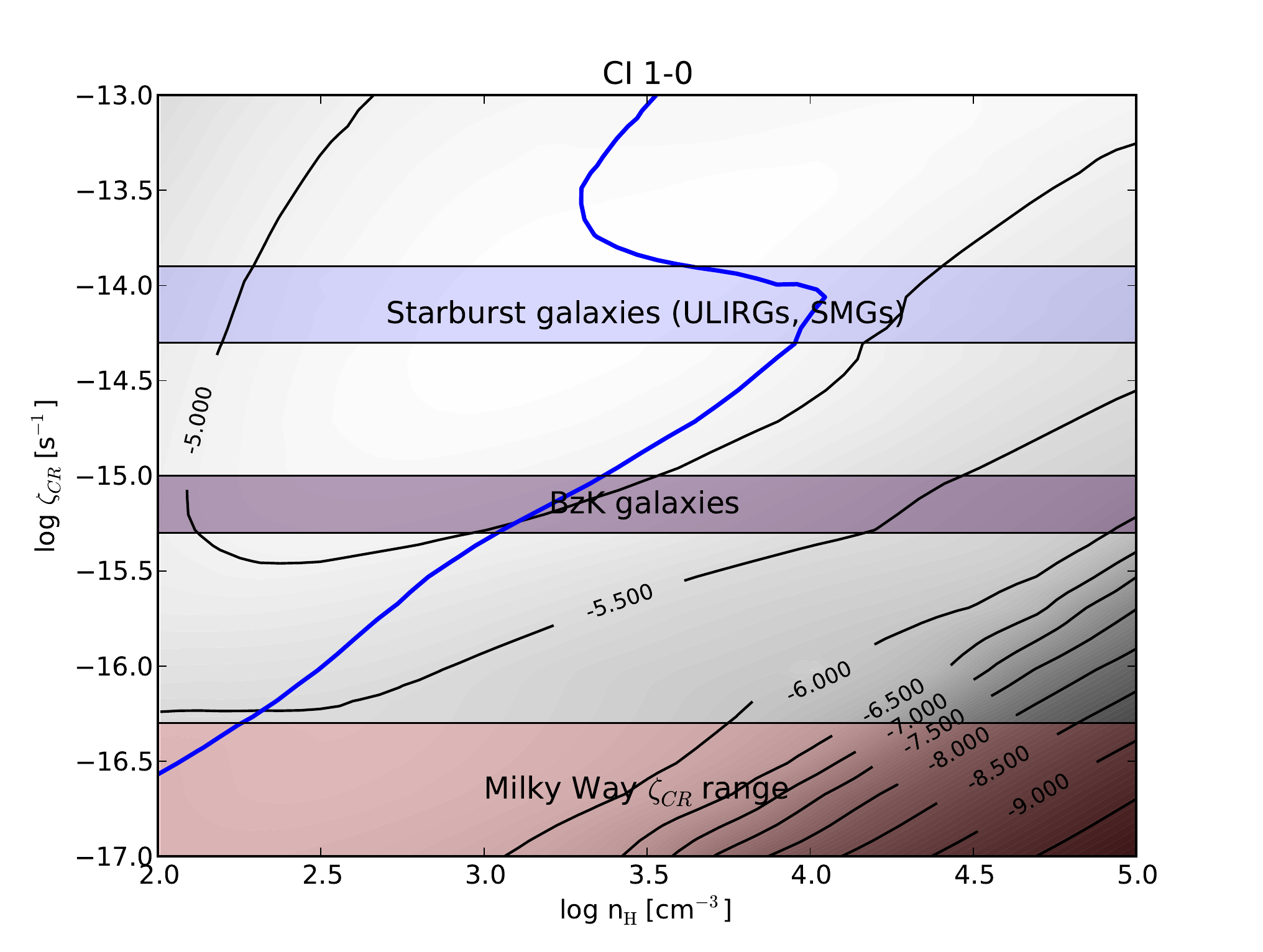}
\includegraphics[width=0.45\textwidth]{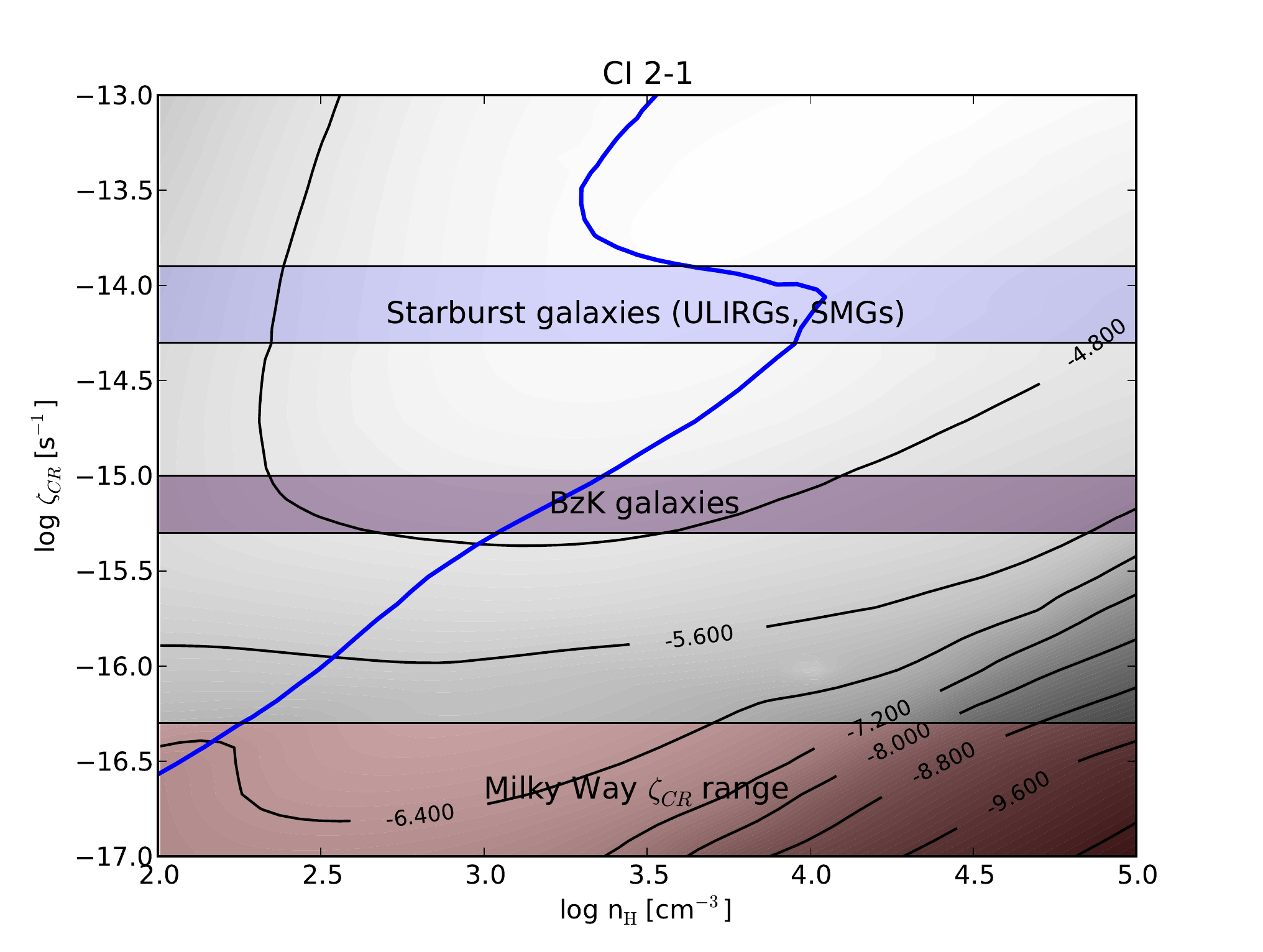}
\caption{ Logaritmic contour/grayscale plots showing  C{\sc i} $J$=1--0 (left) and C{\sc i} $J$=2--1 (right) brightness in the $[n_{\rm H}-\zeta_{\rm CR}]$ space. The brightness is calculated by integrating the local emissivity for a depth of $A_{\rm V}=10\,{\rm mag}$, and is in units of ${\rm erg}\,{\rm cm}^{-2}\,{\rm s}^{-1}\,{\rm sr}^{-1}$. The blue line corresponds to CO/H$_2=10^{-5}$ (see Fig. \ref{fig:zcr} for further details). The two C{\sc i} lines remain luminous for much of the $[n_{\rm H}-\zeta_{\rm CR}]$ space, even in CO poor regions. They become faint (as well as the CO $J$=1--0, 2--1 lines) only for dense/cold gas with Galactic levels of $\zeta_{\rm CR}$ (see \S\ref{ssec:intensity} for detailed discussion).}
\label{fig:CIbright}
\end{figure}

Figure \ref{fig:CObright} plots the line-integrated intensities of the CO $J$=1--0 and $J$=2--1 transitions. It can immediately be seen that these intensities fall precipitously towards the CO-poor area. Indeed, their strengths relative to those typical in the CO-rich regions range from $\sim $0.39 at the ``crossing'' of the blue line (used to mark the CO-rich/CO-poor transition zone in the $\rm [n_H-\zeta _{CR}]$ parameter space), to only $\sim $0.025-0.15 deeper into the CO-poor regions in the upper left corner of the parameter space. Thus CO line observations conducted to yield the same S/N ratio for CO-poor molecular gas distributions would need $\sim $7-1600 longer integration time with respect to those of CO-rich regions. The C{\sc i} $J$=1--0, 2--1 lines on the other hand shown in Fig. \ref{fig:CIbright}, remain bright throughout most of the $\rm [n_H-\zeta _{CR}]$ parameter space, and actually much brighter than the two CO lines\footnote{This does not translate to $\rm I_{C{\sc i}}/I_{CO}$$\times $ higher S/N ratios for C{\sc i} versus CO observations as $\rm T_{sys}(\nu)$ will be higher at C{\sc i} line frequencies. Nevertheless, significant gains in S/N are expected \citep[e.g.][]{Papa04}.} This enables the two C{\sc i} lines to trace $\rm H_2$ gas both in CO-poor and CO-rich regions.

There is also an area of low CO {\it and} C{\sc i} line intensities in the (high-$\rm n_H$)/(low-$\rm \zeta _{CR}$) region of the  $\rm [n_H-\zeta _{CR}]$ parameter space  (lower right corner) in all plots of both Figs. \ref{fig:CObright} and \ref{fig:CIbright}. This is the area where non-SF, dense gas cores  embedded deep into Galactic GMCs are located (e.g. Bok globules). These are very cold regions ($\rm T_{\rm gas}$$\sim $10\,K), and the reduction of CO $J$=1--0, 2--1 line intensities with respect to the warmer gas in the (lower-$\rm n_H$)/(higher-$\rm \zeta _{CR}$) areas almost purely reflects a temperature drop (since CO $J$=1--0, 2--1 are thermalized and optically thick with $\rm E_{10,21}/(k_B T_{\rm gas})\la 1$). The drop of the  C{\sc i} line intensity  towards that area is due  both to this temperature drop (that contributes non-linearly to the intensity reduction since for these lines $\rm E_{10,21}/(k_B T_{\rm gas})\ga 1$, especially for C{\sc i} $J$=2--1), and a [C/CO] abundance reduction. Nevertheless, given that such dense, cold gas cores contain only small mass fractions ($\la $1\%) of Galactic GMCs, we do not anticipate that this (CO and C{\sc i})-dark region of $\rm [n_H-\zeta _{CR}]$ parameter space to be of any consequence when it comes to the inventory of H$_2$ gas mass.

\subsection{The effect of higher FUV fields}

The choice of $G=0G_{\circ}$ in our calculations was driven by our intent to isolate the effect of CRs on the average [CO/H$_2$] abundance in molecular clouds. However ISM environments with higher SFR densities boost both the average $\rm \zeta _{CR}$ {\it and} the average FUV field $\rm G$. Nevertheless, unlike $\rm \zeta _{CR}$ the boost of the average FUV radiation field is much reduced by the dust absorption in the metal-rich ISM environments we study here. Indeed it is expected to be caped to $\rm G\sim 10^3G_{\circ}$ \citep{Papa14}, with higher values such as $\sim 10^{4}-10^{5}$ associated with gas regions near O, B stars that contain only small mass fractions of molecular clouds. Furthermore, in environments of high gas turbulence (typical for ISM with high SFR densities, see also discussion in \S\ref{ssec:caveats}) such clouds are expected to have large $\langle A_V\rangle$, much larger than $\rm \langle A_V \rangle \sim 7-8$ found for Galactic GMCs \citep[see][]{Pelu06}. Thus we do not expect the [CO/H$_2$] to be affected by the average FUV fields for much of the molecular gas mass, a state of affairs which may, however, change for low-metallicity ISM. We must however note that increasing the FUV will result to a higher C{\sc ii} abundance (and C{\sc ii} line emission) in low densities.

We nevertheless still performed calculations for $G=10^3G_{\circ}$ and $G=10^5G_{\circ}$, and found that the corresponding map shown in Fig. \ref{fig:zcr} to be in excellent agreement for $A_{\rm V}\ge7\,{\rm mag}$. In addition, we compare the gas temperature of selected simulations with those studied in \citet{Meij11} who included the effect of FUV radiation and also find excellent agreement.

\subsection{Comparison with other theoretical models, and a [CO/H$_2$] sensitivity on $\rm T_{\rm gas}$}

\begin{figure}[htbp]
  \centering
\includegraphics[width=0.45\textwidth]{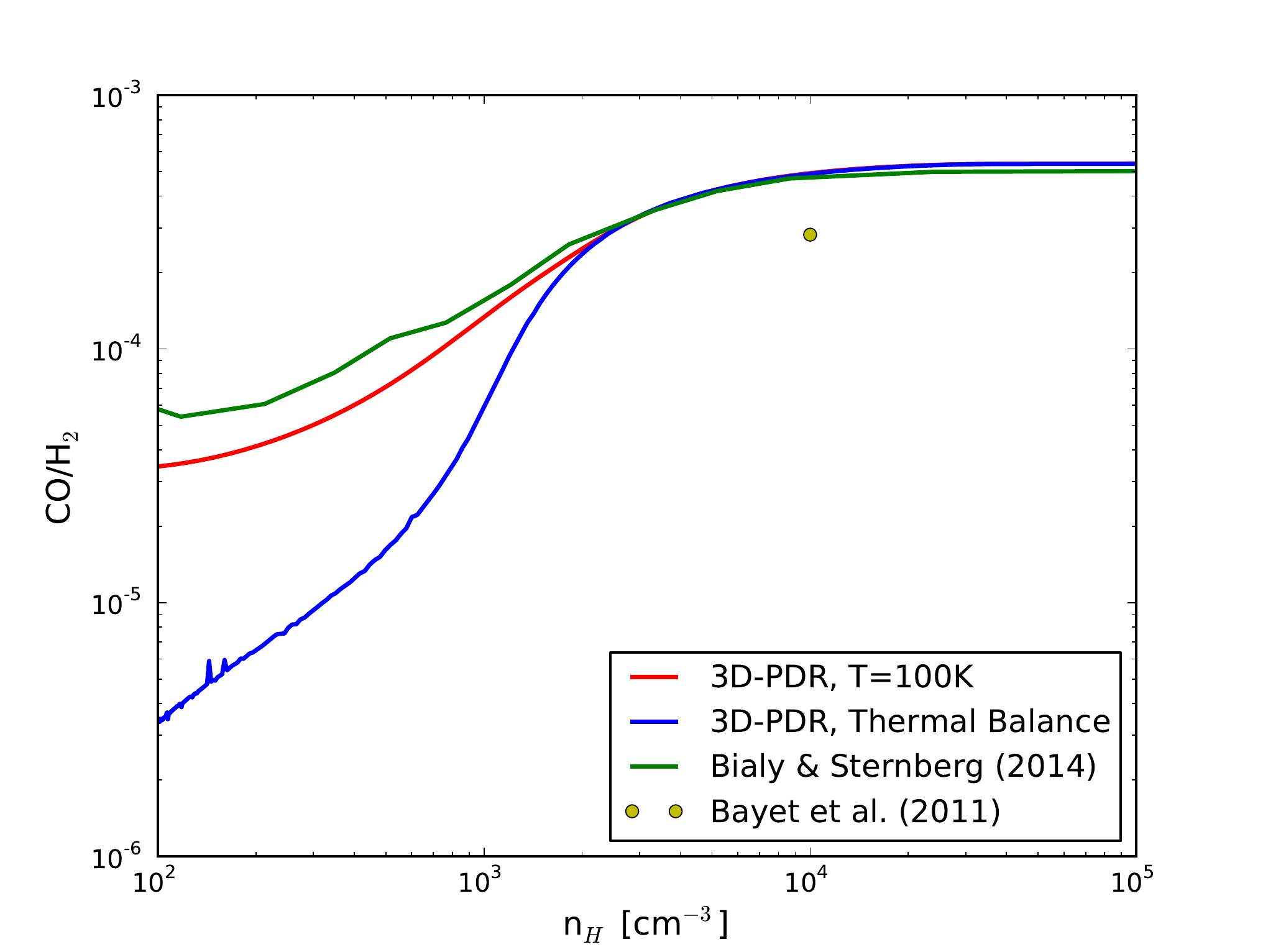}
\caption{ Comparison between {\sc 3d-pdr}, \citet{Baye11,Bial14}. The cosmic-ray ionization rate is $\zeta_{\rm CR}=10^{-16}\,{\rm s}^{-1}$. The green line corresponds to the \citet{Bial14} isothermal simulations and the red line mark our {\sc 3d-pdr} isothermal calculations at $T_{\rm gas}=100\,{\rm K}$. The blue line corresponds to the {\sc 3d-pdr} calculations that solve also for gas thermal balance. The yellow circular point corresponds to a model by \citet{Baye11} for $n_{\rm H}=10^4{\rm cm}^{-3}$. See \S\ref{sec:effectCR} for the relevant discussion.}
\label{fig:comparison}
\end{figure}

As we noted in the Introduction, theoretical models by \citet{Baye11,Meij11,Bial14} have examined the effects of CRs in uniform density clouds, even if they did not focus on the CR-controlled [CO/H$_2$] abundance and its observational consequences in various extragalactic environments. Figure \ref{fig:comparison} shows a comparison between \citet{Baye11,Bial14} and {\sc 3d-pdr} results. Red and green lines compare {\sc 3d-pdr} against \citet{Bial14} respectively for isothermal runs at $T_{\rm gas}=100\,{\rm K}$ interacting with $\zeta_{\rm CR}=10^{-16}\,{\rm s}^{-1}$. It can be seen that the agreement is very good. We note that \citet{Bial14} and the present work used different chemical network which can explain the discrepancy of the two curves for $n_{\rm H}<2\times10^3\,{\rm cm}^{-3}$, while for $n_{\rm H}\ge2\times10^{3}\,{\rm cm}^{-3}$ the agreement is excellent. Switching on the thermal balance calculations (blue line) makes the CO/H$_2$ ratio lower for $n_{\rm H}<2\times10^3\,{\rm cm}^{-3}$ while for $n_{\rm H}\sim10^2\,{\rm cm}^{-3}$ this ratio is $<10^{-5}$, differing by an order of magnitude with respect to the isothermal runs. This demonstrates the importance of thermal balance calculations in estimating the effect of CRs on CO/H$_2$ as the latter appears to have high sensitivity on gas temperature (see also \S\ref{ssec:notb}). Lower temperatures form O$_2$ via the reaction
\begin{eqnarray}
{\rm O} + {\rm OH} \rightarrow {\rm O}_2 + {\rm H}
\end{eqnarray}
which then reacts with C to form CO as shown in Reaction \ref{7} (see \S\ref{ssec:chemistry}), hence increasing the abundance of the latter.

The above reaction rate can be estimated using the following equation \citep[e.g.][]{Mill97}:
\begin{eqnarray}
k=\alpha \left(\frac{T_{\rm gas}}{300}\right)^{\beta} \left(e^{-\gamma/T_{\rm gas}}\right),
\end{eqnarray}
with $\alpha=3.69\times10^{-11}$, $\beta=-0.27$ and $\gamma=12.9$. For $T_{\rm gas}=14\,{\rm K}$ (as obtained by the {\sc 3d-pdr} runs), $k=3.36\times10^{-11}$ while for $T_{\rm gas}=100\,{\rm K}$, $k=4.36\times10^{-11}$. Hence at lower temperatures, such as those found by the thermal balance iterations in this work, the abundance of CO will be higher.

%
%
%

Finally, the yellow circular point corresponds to a simulation presented in \citet{Baye11} who included thermal balance calculations but used different chemical network and initial elemental abundances of species resulting in a discrepancy between their work and the present paper. By using the \cite{Baye11} initial elemental abundances, we obtain excellent agreement.

\subsection{A low-density CO/H$_2$ upwards turnover at high $\bf \rm U_{CR}$  values}

The  CO/H$_2=10^{-5}$ contour  marks  a near-linear  path across  the $[n_{\rm H}-\zeta_{\rm CR}]$ parameter   space   for   up   to   $\zeta_{\rm CR}\sim10^{-14}\,{\rm s}^{-1}$. However, for higher $\zeta_{\rm CR}$ values there seems to be a ``turnover'' of this  CO/H$_2$ contour value  towards lower  densities, i.e.  low-density  gas starts becoming CO-rich  again despite the  now exceptionally high  CR energy densities.  We give a full  description of the underlying chemistry of this behaviour  in \S\ref{ssec:chemistry}. We must  note that although such  CR-intense  environments  are   too  extreme  to  be  considered representatives of average conditions in  the ISM of SF galaxies, they could  be  possible   locally  near  highly  energetic  CR-accelerating phenomena such as supernova explosions and young SNRs.

\subsection{A non-uniform cloud model embedded in a 1G$_{\circ}$ FUV field}
\label{ssec:nonuni}

While our investigation using  uniform density {\sc 3d-pdr} models allows a clear overview of the density-dependence of the CR-induced destruction of  CO, and  what type  of  extragalactic environments  would be  most affected (see Fig.\ref{fig:zcr}),  it is obviously too simple of  an approach when it comes  to  individual  molecular  clouds  where  hierarchical  density structures  are known to  exist.  For  the present  work we  choose to investigate  the   CR-controlled  CO/H$_2$   and  C{\sc i}/CO ratios for a cloud with a simple $n_{\rm H}(r)\propto r^{-1}$ profile. In the future we will use the full capabilities of our {\sc 3d-pdr} code where fractal density structures  have been recently incorporated to examine the CR effects on the CO and C{\sc i} abundances in much more realistic settings \citep[i.e. such as those examined by][]{Walc12}.

We consider a cloud with radius $R=20\,{\rm pc}$ embedded in an environment with different cosmic-ray ionization rates. In particular, we perform a suite of runs in which we use a density profile following the \citet{Lars81} power law of $n_{\rm H}(r)\propto r^{-1}$ by adopting the density function

\begin{eqnarray}
\label{eqn:nhr}
n_{\rm H}(r)=\frac{4\times10^3}{0.04+\left[\frac{r}{{\rm pc}}\right]}\,{\rm cm}^{-3}.
\end{eqnarray}
The above density distribution has the following properties: the central core of the cloud reaches densities up to $n_{\rm H}=10^5\,{\rm cm}^{-3}$ while at $R=20\,{\rm pc}$ it reaches densities down to $n_{\rm H}=200\,{\rm cm}^{-3}$. The total mass of the cloud is $M_{\rm TOT}=2.47\times10^5\,{\rm M}_{\odot}$ while its average density  is $\mean{n_{\rm H}}\simeq300\,{\rm cm}^{-3}$, typical of Galactic GMCs. We also include an external FUV radiation field of intensity $G=1G_{\circ}$. Furthermore, the visual extinction at the centre of the core (measured from the edge of the sphere) is $A_{V, {\rm max}}\sim50\,{\rm mags}$.

\begin{figure*}
  \centering
\includegraphics[width=0.98\textwidth]{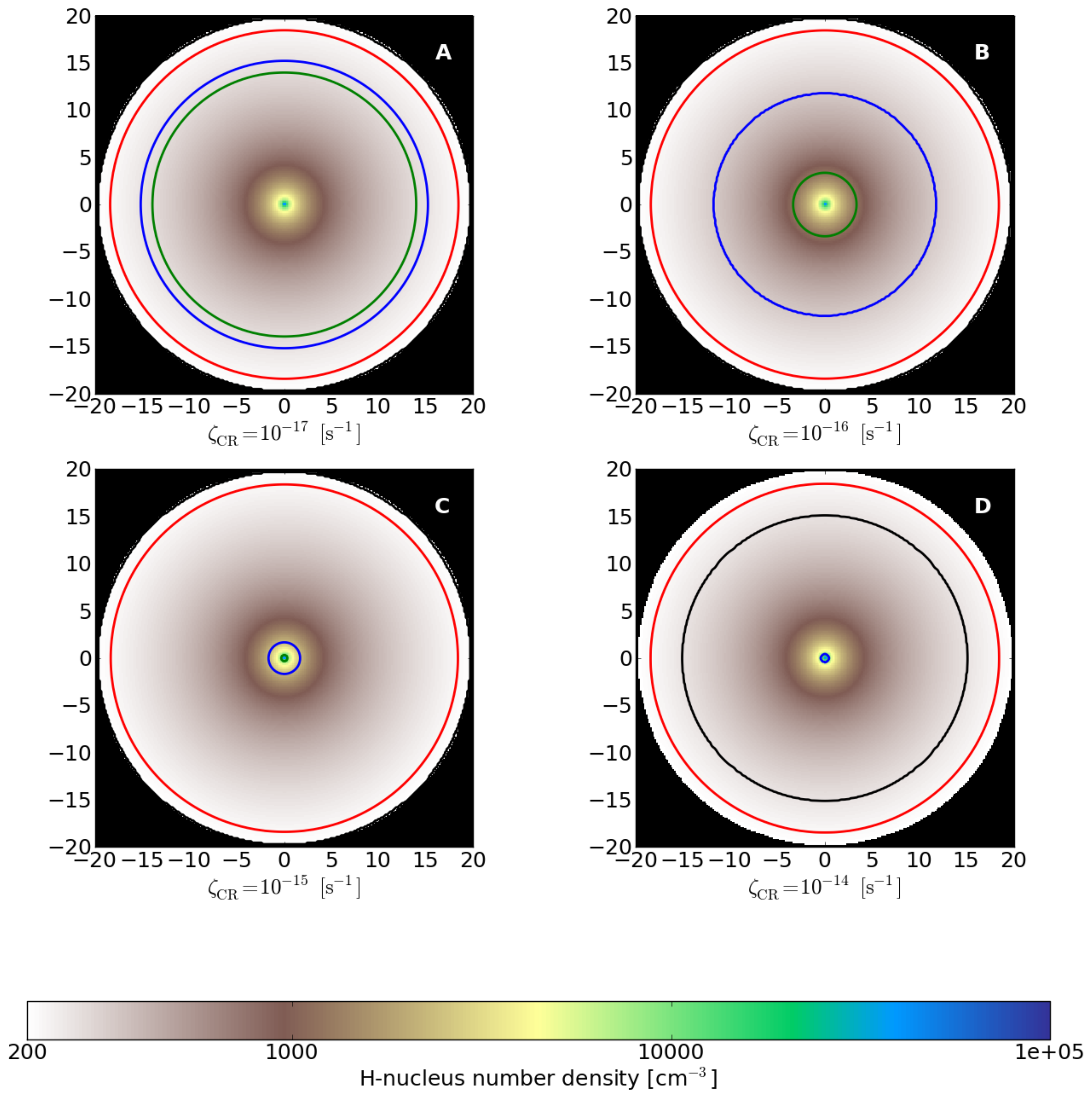}
\caption{ The spherically symmetric one-dimensional simulation for the cloud with density profile as described in Eqn.\ref{eqn:nhr}. Each different panel corresponds to different $\zeta_{\rm CR}$. From A to D, $\zeta_{\rm CR}=10^{-17},\,10^{-16},\,10^{-15},\,10^{-14}\,{\rm s}^{-1}$. The colour bar corresponds to the $n_{\rm H}$ number density. The red line corresponds to C{\sc ii}/C{\sc i}=5. The blue line corresponds to CO/H$_2=10^{-5}$, whereas the green one to C{\sc i}/CO$=5$. The black line (which is only visible in the D panel because in all other cases it lies in the edge of the cloud), corresponds to H{\sc i}/$2$H$_2=1$. H$_2$ is traceable by CO and C{\sc i} in the region below the blue line; by C{\sc i} only in the region between the red and the blue line; and by C{\sc ii} only in the region outside the red line. For the particular case of panel D, H$_2$ is traceable by C{\sc i} only in the inner area of the cloud.}
\label{fig:gmc}
\end{figure*}

Figure \ref{fig:gmc} shows the results of the {\sc 3d-pdr} calculations for different cosmic-ray ionization rates ($\zeta_{\rm CR}=10^{-17} - 10^{-14}\,{\rm s}^{-1}$). We can now see the effects of enhanced CRs within an inhomogeneous $\rm H_2$ cloud. The red line marks a C{\sc ii}/C{\sc i}=5 ratio, with the spherical shell outside that line (i.e. outer part) consisting almost entirely of C{\sc ii}. The green line corresponds to C{\sc i}/CO=5 hence the shell between the red and green lines is C{\sc i}-dominated while the rest of the cloud (i.e. inner to the green line) is CO-dominated. The blue line defines the CO/H$_2=10^{-5}$ ratio and hence the maximum volume where CO can easily trace H$_2$. Finally, we also plot the H{\sc i}/$2$H$_2$=1 ratio with a black line (visible only in Panel D). The volume inner to that line is H$_2$-dominated, with the outer part  where CRs have mostly destroyed H$_2$ to form H{\sc i}.

In Table \ref{tab:abcd} we list the H$_2$ gas mass fractions traceable by CO, C{\sc i} and C{\sc ii} in each case. We find that {\it C{\sc ii} cannot trace more than 1\% of the cloud in any of the cases.} As $\zeta_{\rm CR}$ increases, the H$_2$ gas becomes  C{\sc i}-rich as CO gets destroyed by the reaction with He$^+$. At $\zeta_{\rm CR}=10^{-16}\,{\rm s}^{-1}$ only $\sim33\%$ remains  CO-rich while for $\zeta_{\rm CR}\ge10^{-15}\,{\rm s}^{-1}$ most of the cloud is CO-poor (but C{\sc i}-rich). Evidently, cosmic rays play a major role in controlling the CO-visibility of H$_2$ gas while leaving the C{\sc i} lines as the only alternative molecular gas mass tracers.

\begin{table*}
\begin{center}
\caption{ This table shows the amount (\%) of H$_2$ gas which is traceable by CO, C{\sc i} and C{\sc ii} for the four simulations presented in \S\ref{ssec:nonuni}. The first column corresponds to the panels of Fig.\ref{fig:gmc}. The second column corresponds to the intensity of CRs and the third column to the H$_2$ gas mass. The rest of columns correspond to the amount of H$_2$ that is CO, C{\sc i}, and C{\sc ii} bright i.e. traceable by these elements. It can be seen that once we increase the $\zeta_{\rm CR}$ value C{\sc i} becomes the most important tracer for H$_2$.}
\label{tab:abcd}
\begin{tabular}{cccccc}
Panel & $\zeta_{\rm CR}$ $[\times10^{-17}\,{\rm s}^{-1}]$ & H$_2$ $[\times10^5\,$M$_{\odot}]$ & CO \& C{\sc i} (\%)& C{\sc i} (\%)& C{\sc ii} (\%)  \\
\tableline
A & 1      & 2.47 & $\sim$55 & $\sim$44 & $<$1 \\
B & 10     & 2.47 & $\sim$33 & $\sim$76 & $<$1 \\
C & $10^2$ & 2.47 & $<$1 & $\sim$99 & $<$1 \\
D & $10^3$ & 1.38 & $<$0.1 & $>$99.9 & N/A\\
\tableline
\end{tabular}
\end{center}
\end{table*}

\subsection{The importance of thermal balance}
\label{ssec:notb}

\begin{figure}[htbp]
  \centering
\includegraphics[width=0.45\textwidth]{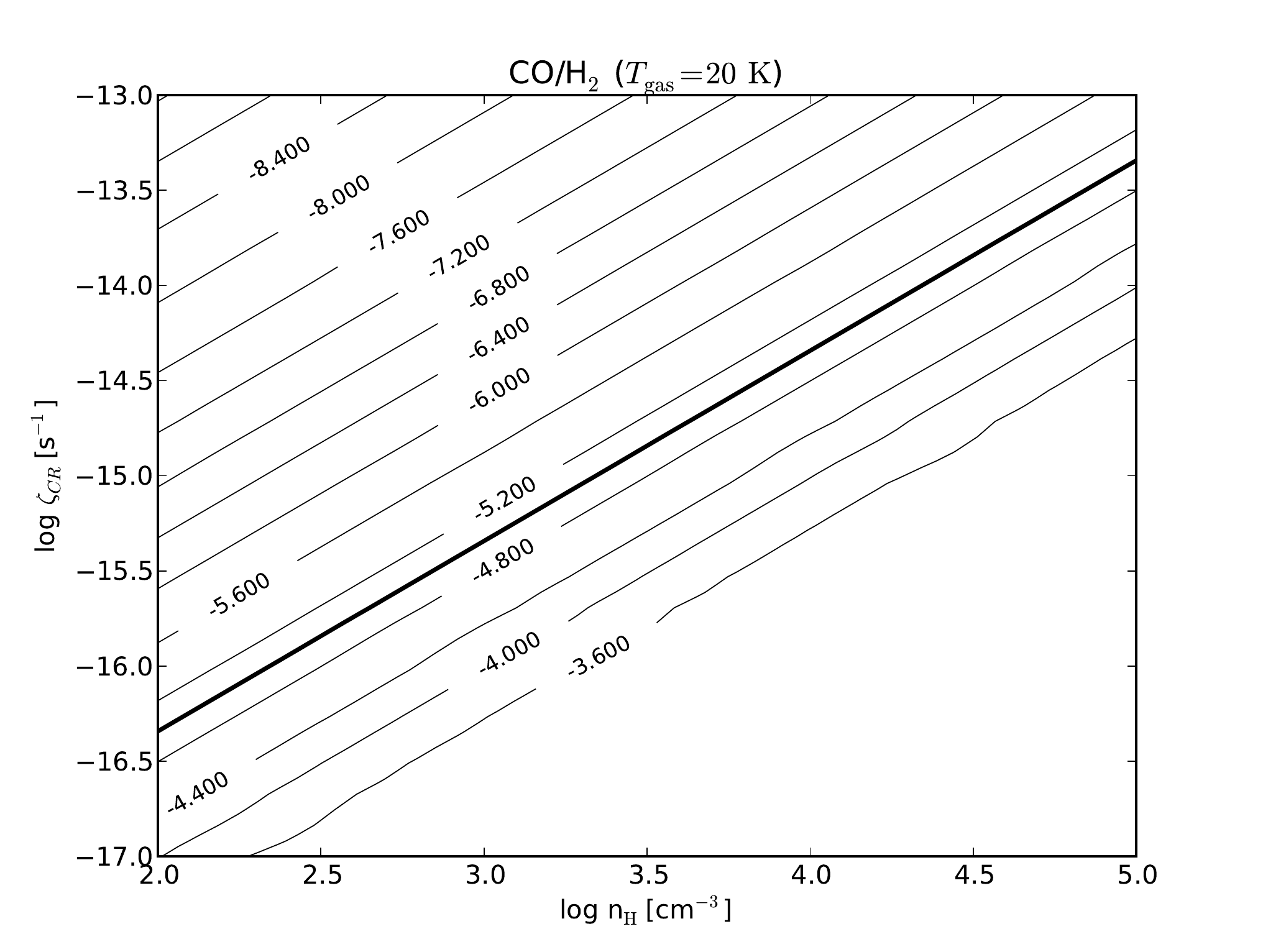}
\includegraphics[width=0.45\textwidth]{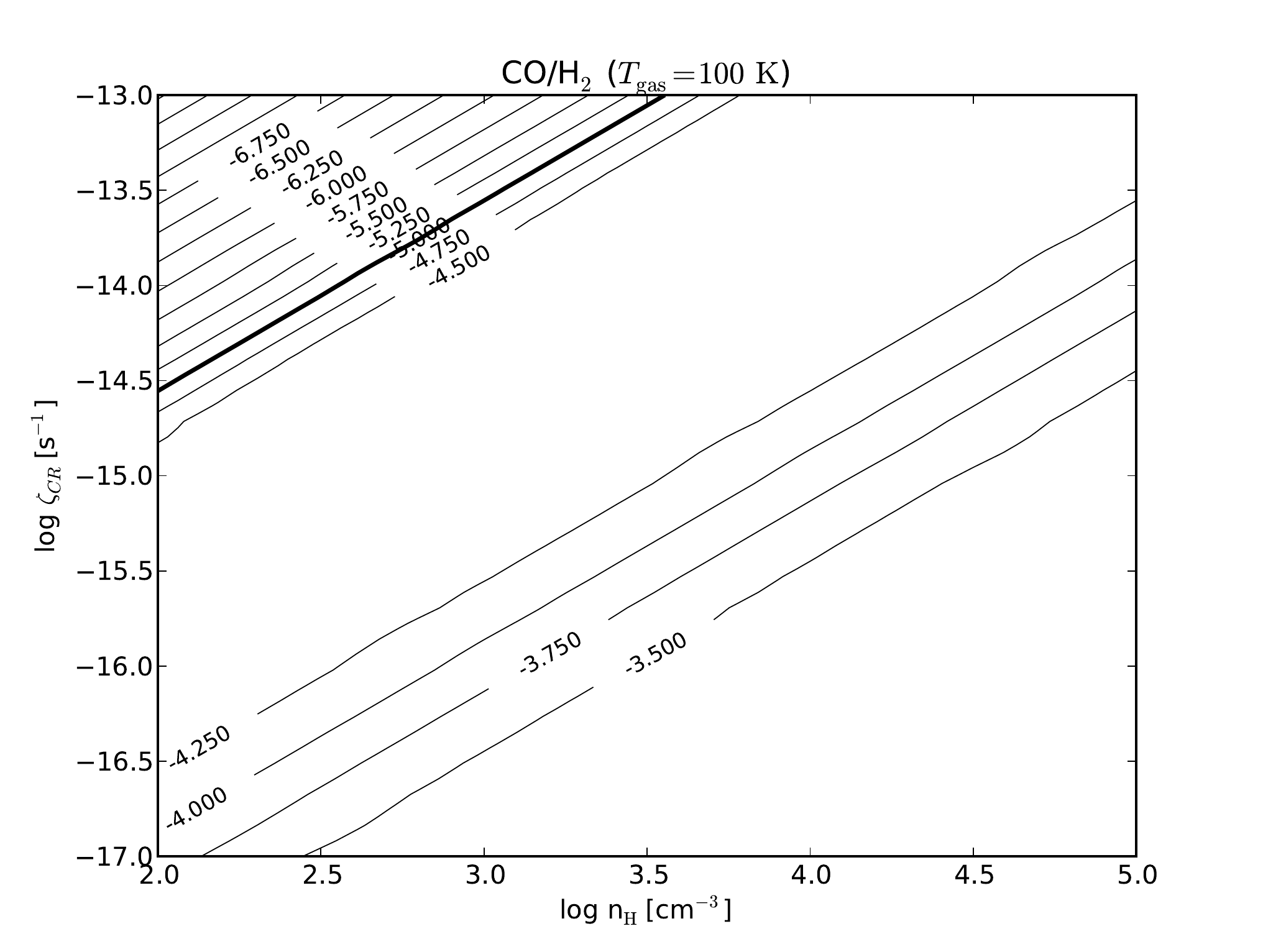}
\caption{ The CO/H$_2$ ratio for the isothermal runs discussed in \S\ref{ssec:notb}. The left panel shows the resultant contour plot for $T_{\rm gas}=20\,{\rm K}$ and the right panel for $T_{\rm gas}=100\,{\rm K}$ \citep[c.f.][]{Bial14}. The thick solid line corresponds to CO/H$_2=10^{-5}$. A strong dependency on the gas temperature is apparent, with the highest temperatures corresponding to much smaller surpressions of the CO/H$_2$ abundance ratio. This demonstrates the importance of solving for the detailed thermal balance. }
\label{fig:notb}
\end{figure}

We will now briefly explore the effects of thermal balance on the results of this work. To do this, we switch off the thermal balance iteration scheme in {\sc 3d-pdr} and instead we run two isothermal grids of runs. Figure \ref{fig:notb} plots the CO/H$_2$ ratio in the $n_{\rm H}$ and $\zeta_{\rm CR}$ parameter space for $T_{\rm gas}=20\,{\rm K}$ (left panel) and $T_{\rm gas}=100\,{\rm K}$ (right panel). The thick solid line corresponds to CO/H$_2=10^{-5}$. In both cases, the CO/H$_2$ ratio follows a linear behaviour with $\log n_{\rm H}$ and $\log \zeta_{\rm CR}$ thus demonstrating that the dependence of the chemical reactions on the gas temperature in such cosmic-ray dominated environments is a highly non-linear and at the same time a very important one. Indeed \citet{Bial14}, which used an analytical approach (and simpler chemical pathways) along with an isothermal assumption (at $T_{\rm gas}=100\,{\rm K}$), do not recover this behavior, nor do they find the large reduction of CO/H$_2$ at high CR energy densities.

\section{A  secondary effect: clumpy low-$J$ CO emission in gas-rich SF disks?}

A secondary  effect that may occur in H$_2$-rich galactic disks  permeated  by high $\rm  U_{CR}$  backgrounds (boosted  by  the  intense star  formation typically hosted by such disks)  which destroy CO more effectively  at lower gas densities {\it is to make such disks appear clumpy in low-$J$ CO line emission}\footnote{Molecular  gas disks   in CO  $J=3-2$ and higher-$J$ CO line emission {\it are} expected to appear much clumpier than in CO $J=1-0$, $2-1$    because high-$J$ CO lines sample  only the more compact dense and warm  gas regions near SF sites while CO $J$=1--0,    2--1 are expected to trace the entire molecular gas distribution  (as long  as it is CO-rich).}.  This can occur     simply because the  CR-induced destruction of CO, being more effective at lower  densities, will leave only      smaller and denser parts of molecular clouds  to be CO-rich. The appearance of such clouds or cloud ensembles       via their low-$J$ CO line emission will then be much  clumpier, depending on the average $\rm  U_{CR}$  background        and  the underlying  mass-density  $\rm dM_{gas}(n)/dn$ function of the molecular clouds. Finally, if lower density        molecular clouds lie  preferably  at  larger  galactocentric  distances,  such CR-inundated  molecular gas disks        could appear more  compact in low-$J$ CO emission than they actually are.   Such large scale CR-induced        effects on the (CO-line)-visibility of the H$_2$ gas distribution in galaxies can only be accentuated once        metallicity gradients (with lower-metallicity gas found preferably at larger galactocentric distances) are also taken        into account (see also \S\ref{ssec:caveats}).

There  are now  several  deep  CO $J=1-0$, $2-1$  images available  for gas-rich SF systems in the distant Universe \citep[e.g.][]{Dadd10, Ivis10, Riec11, Hodg12}, with clumpy  gas  disks  often   claimed \citep[e.g.][]{Hodg12}. In such cases it would be interesting to also conduct deep C{\sc i} $J$=1--0, 2--1 line imaging since C{\sc i} is a by-product of the CR-induced CO destruction which remains abundant also in CO-rich molecular gas. Thus it can then trace  any CO-dark $\rm H_2$ gas along with its CO-rich  phase (see Fig.\ref{fig:zcr}), and may thus reveal smoother H$_2$ gas distributions in CR-permeated SF disks than the  CO $J$=1--0, 2--1 lines.

On the theoretical front, hydrodynamical simulations of  gas-rich disks with: a) high enough resolution to track the molecular cloud turbulence (which  sets  their $\rm dM_{gas}(n)/dn$) and b) detailed H$_2$,  CO formation/destruction chemical networks, can  elucidate  whether CRs can induce such a bias on the structural characteristics  of H$_2$-rich disks  derived from  CO $J$=1--0, 2--1 lines (see also \S\ref{ssec:tests}).

\section{A chemical and analytical determination of the CO-dark H$_2$ domain}

A main result of this study is the discovery of a large area in the $[n_{\rm H}-\zeta_{\rm CR}]$ parameter
space where gas is mostly molecular but CO-poor. In this Section we  identify the responsible chemical pathways for this behaviour, then attempt to find an analytical equation that could be used in purely (hydro-)dynamical codes for determining the average CO/H$_2$ ratio as a function of gas density and $\rm \zeta_{CR}$ in turbulent molecular gas clouds.

\subsection{The chemistry of H$_2$ and CO}
\label{ssec:chemistry}

While the {\it main} chemical formation and destruction routes of both H$_2$ and CO in the gas phase are very well known, in order to understand the region in the $[n_{\rm H}-\zeta_{\rm CR}]$ plane of Fig.\ref{fig:zcr} where CO is destroyed but H$_2$ is not, one needs to take  into consideration several more chemical reactions as a function of these two parameters. We perform  a  chemical analysis of three representative such models outlining that region (their properties in tabulated in Table \ref{tab:models} and the rates used for the reactions presented below are shown in Table \ref{tab:rates}). We note, however, that the chemical pathway and particularly the efficiency of the reaction rates follows a non-linear behavior, therefore the analysis below is very model-dependent and hence should be taken as qualitative.

\begin{table}
\begin{center}
\caption{Properties of three representative models marking  the outlines of CO-poor/H$_2$-rich region.}
\label{tab:models}
\begin{tabular}{ccc}
Model ID & $\log\,n_{\rm H}\,({\rm cm}^{-3})$ & $\log\,\zeta_{\rm CR}\,({\rm s}^{-1})$ \\
\tableline
M1 & 2.5 & -16 \\
M2 & 2.5 & -15 \\
M3 & 3.5 & -15 \\
\hline
\end{tabular}
\end{center}
\end{table}

\begin{table}[h]
\begin{center}
\caption{Reaction rates for the chemical analysis discussed in \S\ref{ssec:chemistry}.}
\label{tab:rates}
\begin{tabular}{cccc}
Reaction & $\alpha$ & $\beta$ & $\gamma$ \\
\tableline
\ref{1} & $6.40\times10^{-10}$ & $0.0$ & $0.0$ \\
\ref{2} & $7.98\times10^{-10}$ & $-0.16$ & $1.4$ \\
\ref{2b} & $2.00\times10^{-9}$ & $0.0$ & $0.0$ \\
\ref{2c} & $1.36\times10^{-9}$ & $0.0$ & $0.0$ \\
\ref{3} & $1.20\times10^{-17}$ & $0.0$ & $0.0$ \\
\ref{4} & $2.08\times10^{-9}$ & $0.0$ & $0.0$ \\
\ref{5} & $6.02\times10^{-11}$ & $0.1$ & $-4.5$ \\
\ref{6} & $1.00\times10^{-10}$ & $0.0$ & $0.0$ \\
\ref{7} & $5.56\times10^{-11}$ & $0.41$ & $-26.9$ \\
\ref{9} & $7.50\times10^{-10}$ & $0.0$ & $0.0$ \\
\ref{10} & $1.60\times10^{-9}$ & $0.0$ & $0.0$ \\
\hline
\end{tabular}
\end{center}
\end{table}

In M1 and M2, H$_2$ is mainly formed by the ion-neutral reaction:
\begin{eqnarray}\label{1}
 {\rm H} + {\rm H}_2^+ \rightarrow {\rm H}_2 + {\rm H}^+
\end{eqnarray}
which contributes $\sim$(70-90) \% to the formation of H$_2$.
For the formation of H$_2$ the behaviour is similar for the model M3, with the difference that the ion-neutral reactions forming H$_2$ are:
\begin{eqnarray}\label{2}
{\rm H}_3^+ + {\rm O} \rightarrow {\rm OH}^+ + {\rm H}_2 \\
{\rm H}_3^+ + {\rm C} \rightarrow {\rm CH}^+ + {\rm H}_2 \label{2b}\\
{\rm H}_3^+ + {\rm CO}  \rightarrow {\rm HCO}^+ + {\rm H}_2 \label{2c}
\end{eqnarray}
H$_2$ is then destroyed {\it in almost equal efficiencies} by recombination and cosmic rays ionization in all models:
\begin{eqnarray}
{\rm H}_2 + {\rm c.r.} \rightarrow {\rm H}_2^+ + e^- \label{3} \\
{\rm H}_2^+ + {\rm H}_2 \rightarrow {\rm H}_3^+ + {\rm H} \label{4}
\end{eqnarray}
Hence cosmic rays, although an efficient mean of destruction, are not necessarily the dominant route. CO on the other hand is formed mainly via neutral-neutral reactions in M1, with the two most efficient ($\sim30\%$ each) routes being:
\begin{eqnarray}
{\rm CH} + {\rm O} \rightarrow {\rm CO} + {\rm H}  \label{5} \\
{\rm C} + {\rm OH} \rightarrow {\rm CO} + {\rm H} \label{6}
\end{eqnarray}
with some contribution ($\sim$ 15\%) from:
\begin{eqnarray}
{\rm C} + {\rm O}_2 \rightarrow {\rm CO} + {\rm O} \label{7}
\end{eqnarray}
In M2, the Reaction \ref{7} is not important but the rest of CO is formed via the ion-neutral reaction:
\begin{eqnarray}\label{9}
{\rm H} + {\rm CO}^+ \rightarrow {\rm CO} + {\rm H}^+
\end{eqnarray}
Interestingly, in M3 CO forms only via the neutral-neutral reactions listed above (i.e Reactions \ref{5}--\ref{7}).

However, {\it in all models {\rm CO} is always very efficiently destroyed indirectly by cosmic rays, via He$^+$} (which forms via cosmic ray ionization) with the reaction:
\begin{eqnarray}\label{10}
{\rm He}^+ + {\rm CO} \rightarrow {\rm O} + {\rm C}^+ + {\rm He}
\end{eqnarray}
It  therefore seems that a  main reason for a CO-dark H$_2$ phase is that a higher cosmic ray ionization rate  drives a higher rate of destruction of CO, but has a much less dramatic effect on H$_2$. It is worth noting here that He$^+$ can be destroyed by reacting with H$_2$; this reaction has a small barrier leading to its rate coefficient becoming a factor of two higher at $100\,{\rm K}$ \citep[i.e. the gas temperature that][considered]{Bial14} than at $40\,{\rm K}$ (i.e. the gas temperature of the M3 model), possibly reducing He$^+$ enough to allow CO to survive at higher abundances. A thorough investigation of the chemistry, as a function of temperature, that can lead to a CO-poor/H$_2$-rich environment is beyond the scope of this paper but it is clear that chemistry resulting from a high cosmic ray ionization rate when thermal balance is taken into account is not easily determined a priori. 

\begin{figure}[htbp]
  \centering
\includegraphics[width=0.45\textwidth]{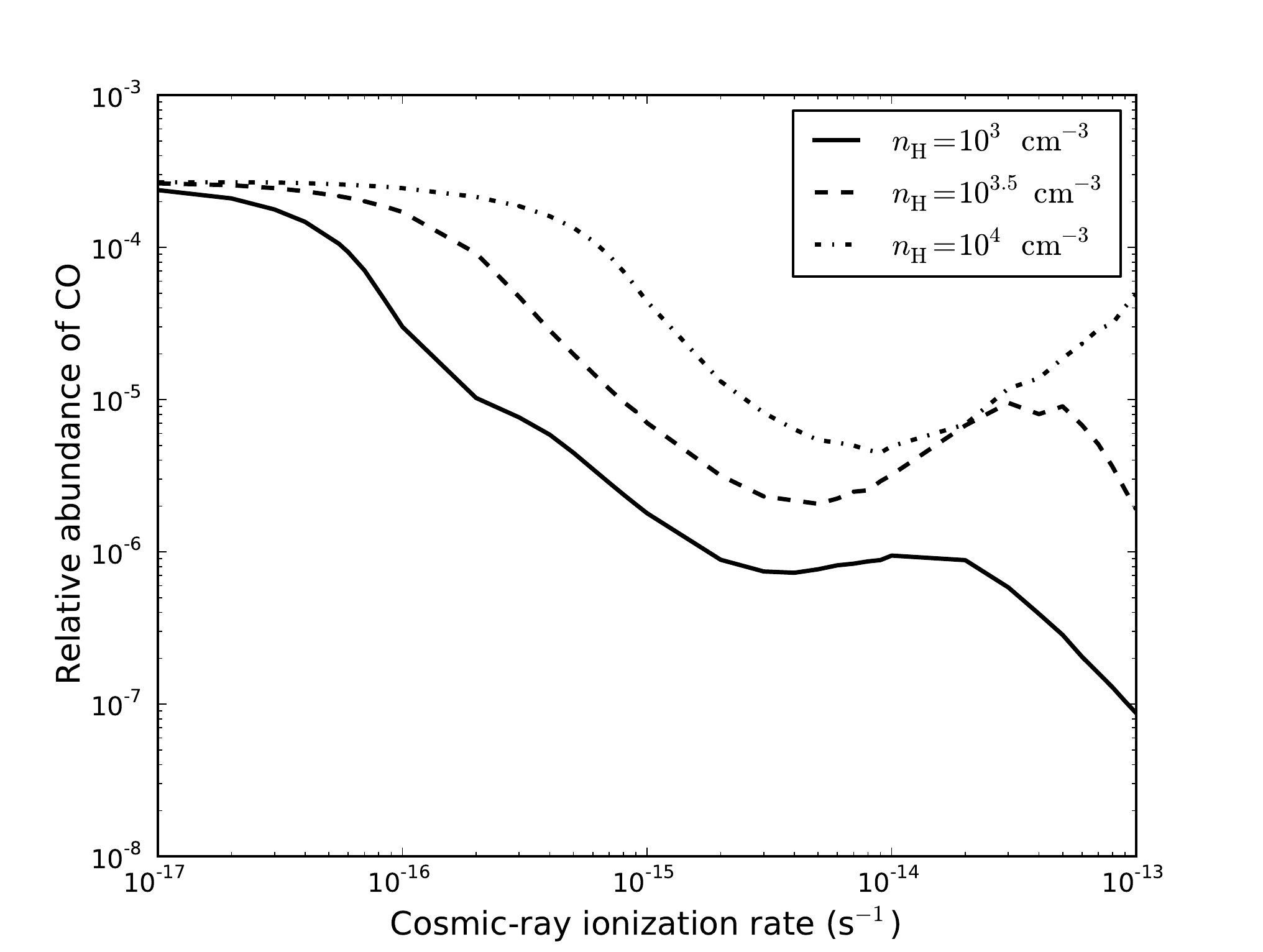}
\includegraphics[width=0.45\textwidth]{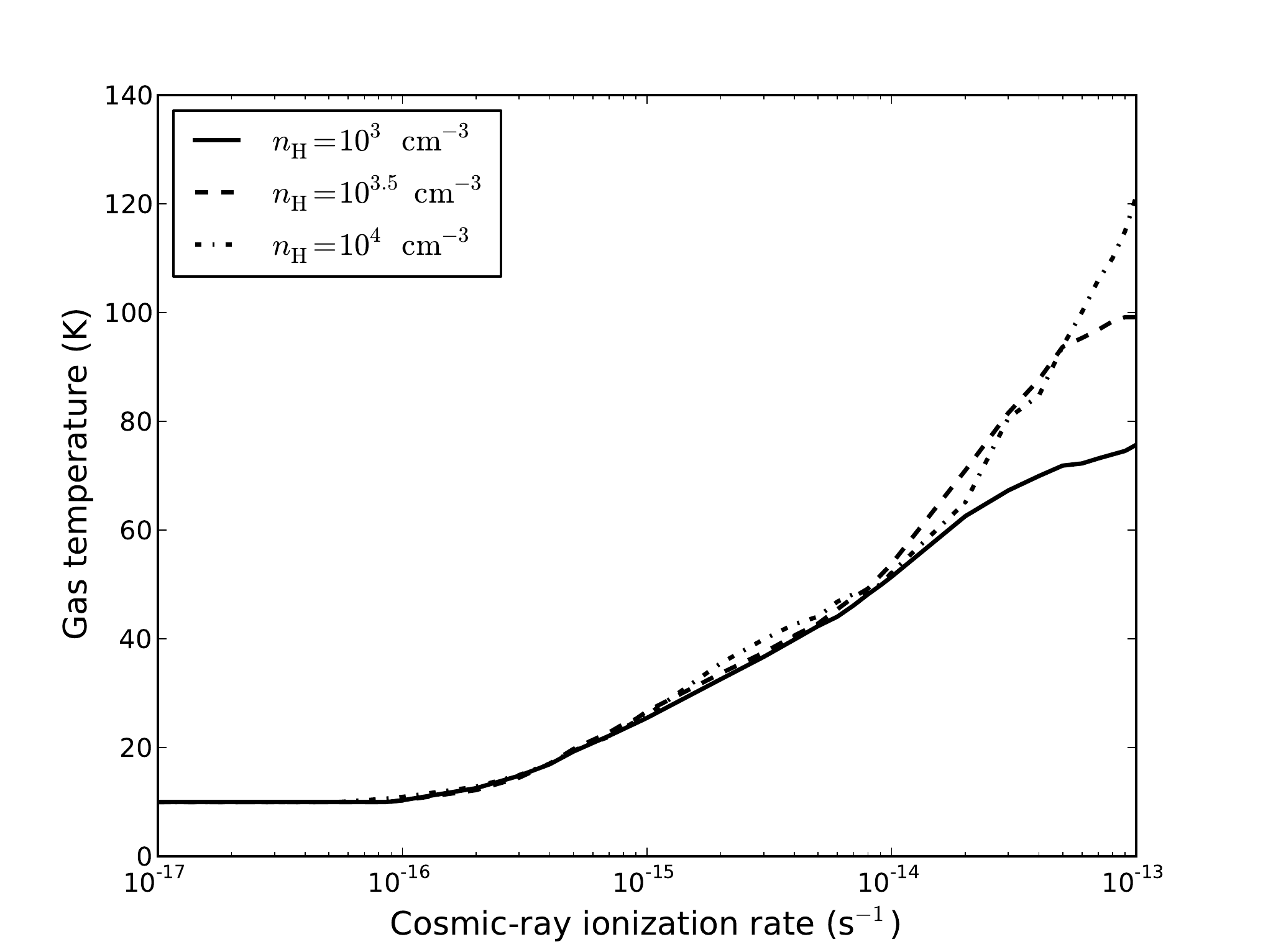}
\caption{ The left panel shows the abundance of CO (relative to hydrogen) versus the cosmic-ray ionization rate ($\zeta_{\rm CR}$) for three different densities ($n_{\rm H}=10^3\,{\rm cm}^{-3}$ solid line, $n_{\rm H}=10^{3.5}\,{\rm cm}^{-3}$ dashed line, and $n_{\rm H}=10^4\,{\rm cm}^{-3}$ dot-dashed line). For $\zeta_{\rm CR}\gtrsim10^{-14}\,{\rm s}^{-1}$ we observe an increase of CO abundance due to the destruction of He$^+$ (see \S\ref{ssec:chemistry}). The right panel shows the corresponding gas temperature which increases with increasing $\zeta_{\rm CR}$.}
\label{fig:xsec}
\end{figure}

 Fig.\ref{fig:zcr} also shows a high $\rm \zeta_{CR}$ regime with a small parameter space where the general trend of (higher-$\rm \zeta_{CR}$)$\rightarrow $(higher-n gas becoming CO-poor) is reversed. In the left panel of Fig.\ref{fig:xsec} we now show how CO/H$_2$ varies with increasing $\zeta_{\rm CR}$ for three different number densities, while the right panel shows the corresponding gas temperatures. The left panel of Fig.\ref{fig:xsec}, for a density range $n_{\rm H}=10^3-10^4\,{\rm cm}^{-3}$ and $\zeta_{\rm CR}\geq 3\times10^{-15}{\rm s}^{-1}$, shows the reversal of the aforementioned trend, with high-density gas becoming CO-rich again despite the now very high $\rm \zeta _{CR}$ values. This reversal is much less prominent for low-density gas.  We  ran a chemical analysis for these models and we found that He$^+$ is efficiently (by $>50\%$) destroyed by:
\begin{eqnarray}\label{11}
{\rm He}^+ + {\rm M} \rightarrow {\rm M}^+ + {\rm He}
\end{eqnarray}
where M stands for metals for which we used a total elemental fractional abundance of 4$\times$10$^{-5}$ \citep{Aspl09}.
Note that Reaction \ref{11} is not very efficient in models where CO decreases with cosmic ray ionization rate.

Nevertheless such  extreme CR-permeated gas regions where  this effect can take place are unlikely to  be representative of  average ISM conditions even  in the most extreme merger/starbursts known, yet  they could be found near or within  young SNRs.  In  such a case  the effect described  here may indicate the survival of the  CO molecule in dense molecular gas  within  some   truly  extreme  CR-dominated  regions.

\subsection{An analytical approximation of the the CO-dark H$_2$ domain}

\begin{figure}[htbp]
  \centering
\includegraphics[width=0.45\textwidth]{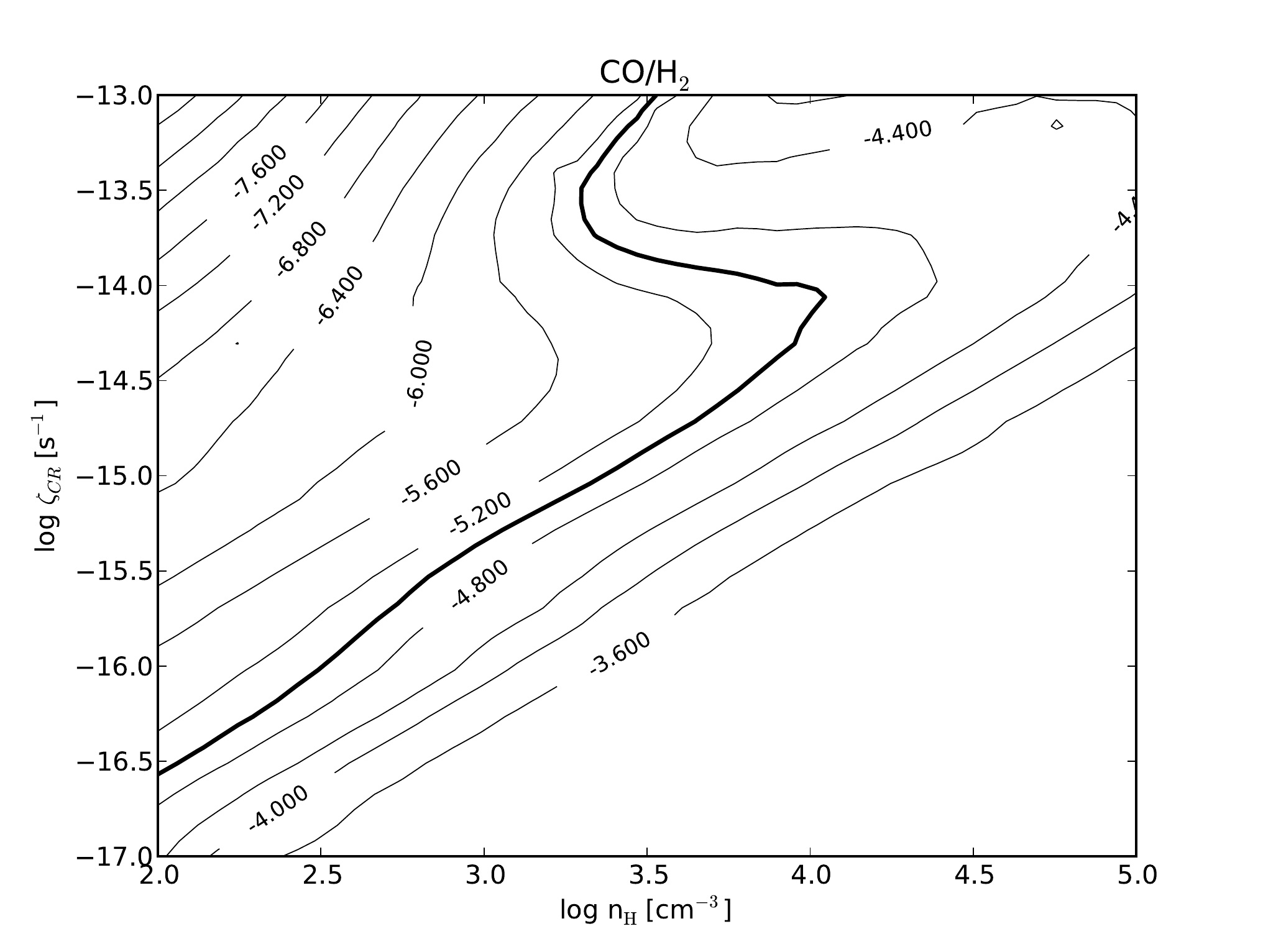}
\includegraphics[width=0.45\textwidth]{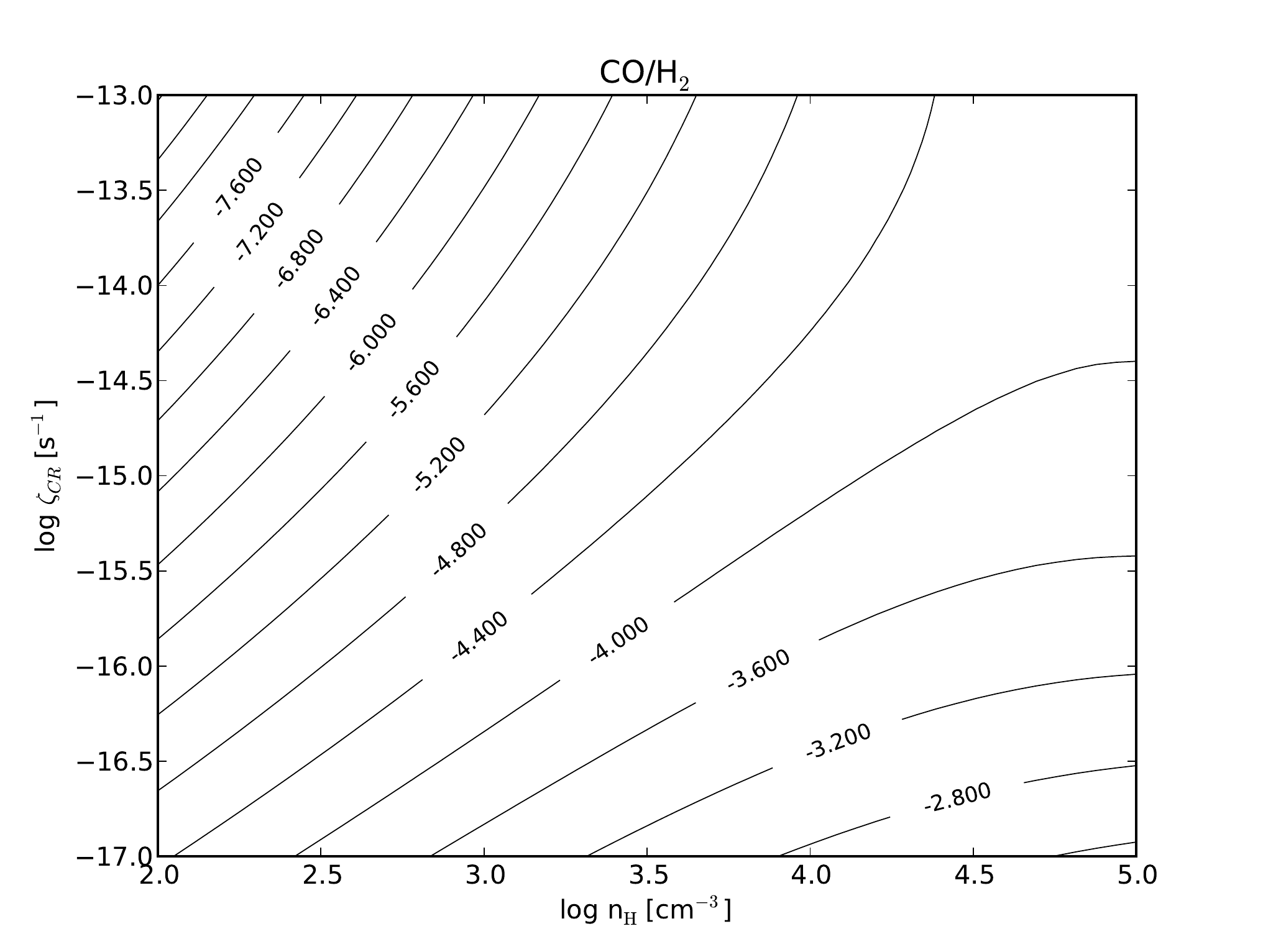}
\caption{ The left panel of the figure shows a contour plot of the {\sc 3d-pdr} results for the CO/H$_2$ ratio. The thick solid line corresponds to CO/H$_2=10^{-5}$. The right panel shows the corresponding best fit function described in Eqn.\ref{eqn:function}.}
\label{fig:coh2}
\end{figure}

The left panel of Fig.\ref{fig:coh2} is a contour plot showing {\sc 3d-pdr} results of the $\log({\rm CO}/{\rm H}_2)$ as a function of $n_{\rm H}$ and $\zeta_{\rm CR}$. The following function is best fit to our data:
\begin{eqnarray}
\label{eqn:function}
&\log({\rm CO}/{\rm H}_2)=-110.7 +28.69\log n_{\rm H} \nonumber \\ \nonumber
&-12.82\log\zeta_{\rm CR} -1.512(\log n_{\rm H})^2 \\ \nonumber
& +2.642\log n_{\rm H}\log\zeta_{\rm CR} -0.5377(\log\zeta_{\rm CR})^2 \\ \nonumber
&-0.08047(\log n_{\rm H})^2\log\zeta_{\rm CR}\\ 
& +0.06209\log n_{\rm H}(\log\zeta_{\rm CR})^2-0.008312(\log\zeta_{\rm CR})^3
\end{eqnarray}
and is plotted in the right panel of Fig.\ref{fig:coh2}. This equation is applicable for negligible FUV radiation within a cloud, typically for $A_{\rm V}\ge7\,{\rm mag}$ (the case for most H$_2$ gas mass in starbursts). In cases where FUV radiation is significant this parametric equation still provides a {\it maximum} CO/H$_2$ abundance ratio possible for a given average SFR density (since FUV can only destroy CO further while leaving the mostly self-shielding H$_2$ intact).

This relation  cannot track the  behaviour of CO/H$_2$  as extracted from  the detailed  {\sc  3d-pdr}  models in  some  domains (e.g.   at $n_{\rm H}\sim1000\,{\rm cm}^{-3}$    and     very    high    cosmic    ray ionization). However for the $[n_{\rm H}-\zeta_{\rm CR}]$ parameter space where most of the molecular gas  mass of SF galaxies is expected to be this relation gives  reasonably good results\footnote{The coefficients   have  a 95\%  confidence  bounds,  and a  Mean  Squared Error  (MSE)  analysis  on  this function  yields  an MSE  of  0.04;  and as  Mean   Absolute Percentage error (MAPE) of  3.5\%.}.  It can then be used in galaxy-sized or cosmology-sized hydrodynamic simulations of atomic and molecular gas  and stars where the underlying  molecular cloud physicsis treated in a sub-grid fashion \citep{Pelu09,Nara11,Lago12,Chri12}.  Such simulations are  frequently used to predict the CO-deduced versus the true  H$_2$ gas distributions  in galaxies \citep[e.g.][]{Pelu09} or  make predictions  of CO  line  luminosities of galaxies in a cosmological setting \citep[e.g.][]{Lago12}. Yet none currently  takes into  account  the expected  strong  effects of  CRs, driven by the often high  and strongly evolving SFR densities found in the Early  Universe.

\begin{figure}[htbp]
  \centering
\includegraphics[width=0.45\textwidth]{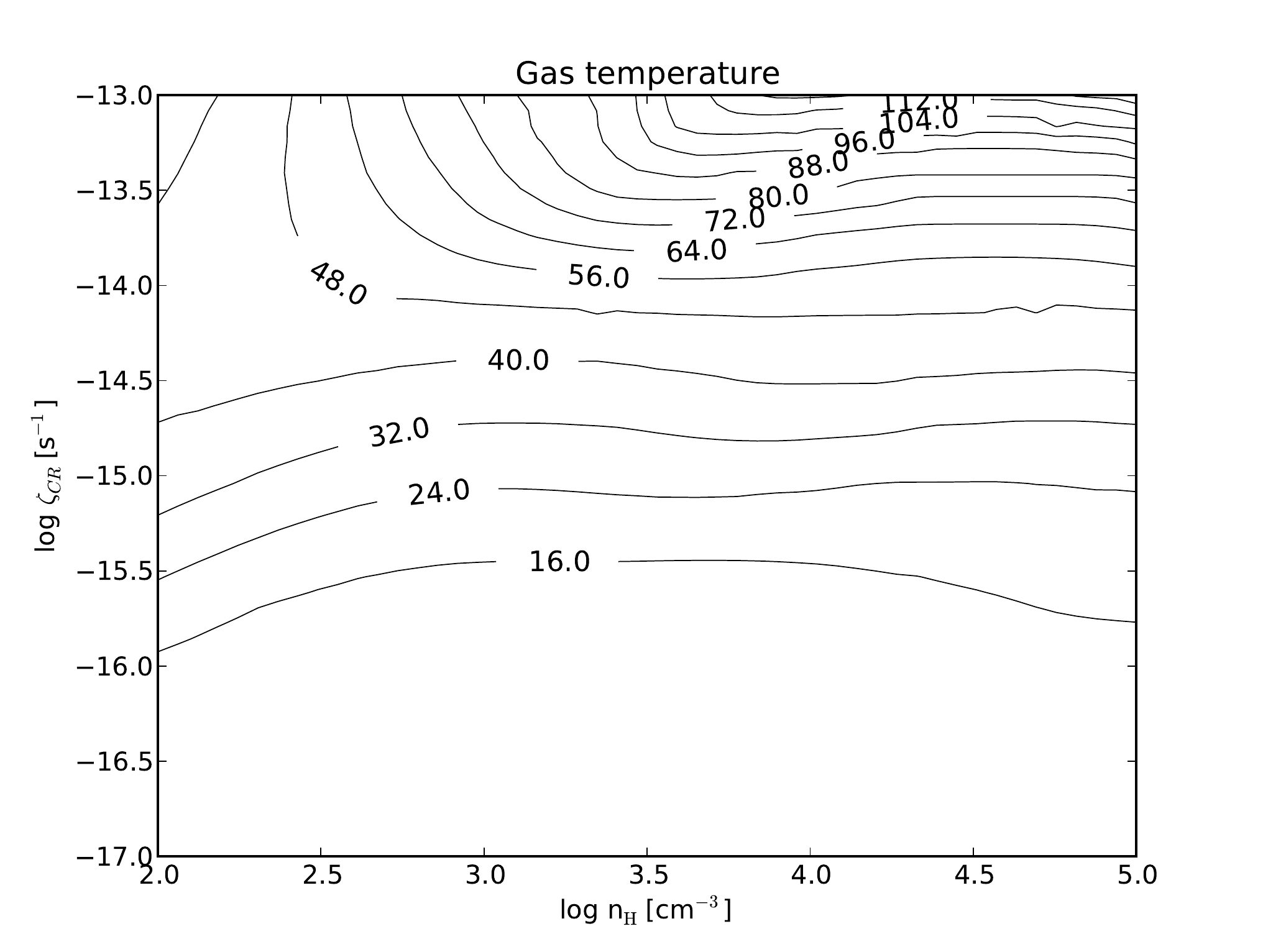}
\caption{ Contours showing the corresponding gas temperatures for the parameter space shown in Fig.\ref{fig:zcr}. From this Figure is also obvious that most of the temperatures attained by the gas are well below the temperature assumed by \citet{Bial14} their isothermal treatment.}
\label{fig:tgas}
\end{figure}

In the case  of cosmological simulations used to  predict CO SLEDs for galaxies in a  $\Lambda$CDM universe \citep[e.g.][]{Lago12} knowing the temperature of CR-heated gas  in the CO-rich and CO-poor phases is also  important.   In  Fig.\ref{fig:tgas}  we  show  the  temperatures attained  by the  molecular  gas within  the  $[n_{\rm H}-\zeta_{\rm CR}]$ parameter space  explored. Finally, from  our work is now clear that   C{\sc i}  $J$=1--0,  2--1 line emission is important  for recovering the  CO-poor H$_2$ gas distribution along  with the CO-rich one  in galaxies.   Cosmological simulations  of C{\sc i}  line luminosities in galaxies have been recently published \citep{Toma14}, but  without taking  into  account  the important  effects  of CRs  in regulating  the CO/H$_2$  and C{\sc i}/H$_2$   abundance ratios.

\section{Discussion}

Quite unlike  the FUV radiation  fields incident on  molecular clouds, CRs can destroy CO  equally efficiently throughout their volume rather than mostly at outer cloud layers.  CRs can thus be the main regulator of the CO-$''$visibility$''$ of molecular  gas in SF galaxies with the FUV-induced CO destruction a  secondary one.  Only for low-metallicity molecular   gas  in  high-intensity   radiation  the   FUV-induced  CO destruction can  surpass the CR-induced one.  Nevertheless, given that high average FUV fields are found in places of elevated SFR densities, and  thus also of boosted  $\rm  U_{CR}$, the  two  effects will  always operate in parallel, even as the average FUV radiation fields will not follow the rise of $\rm \zeta _{CR}$ to high values in a proportional manner in metal-rich SF environments \citep[see][for details]{Papa14}.

In  future work we will explore  the combined effects of high $\rm G_{\circ}$ and $\rm \zeta _{CR}$ on  molecular gas with a range of metallicities. Given the non-linear  effects behind the FUV-induced and  CR-induced CO destruction, these may be much more dramatic  than either  effect alone.   FUV-intense environments with low average metallicities  will be typical in the early  Universe, either over entire galaxies (e.g.   Ly-break  galaxies, dwarfs) or in the outer,  metal-poor,  parts of SF disks  in  the  process  of  building  up  their  present-epoch  Z(R) gradients.

\subsection{Two caveats: CR penetration, and the effects of turbulence}
\label{ssec:caveats}

Our analysis has been based  on the assumption that $\rm U_{CR}\propto \rho_{SFR}  $, and  that  CRs penetrate  into  most of  the volume  of molecular  clouds.   There  are  good  observational  and  theoretical arguments supporting both  these assumptions, reviewed in \citet{Papa10a}.    Briefly,  the  first   assumption  underlies   the  so-called far-IR/radio  correlation  established for  galaxies  whose radio  and far-IR emission are powered by star formation, while the second one is supported by $\gamma$-ray  observations of $\rm H_2$-rich starbursting galactic centers,  and by the thermal, ionization,  and chemical states observed for FUV-shielded dense molecular gas cores in the Galaxy \citep{Berg07}. Finally, if CRs were unable to penetrate deep inside the FUV-shielded regions of molecular clouds there would be no molecules present other than $\rm H_2$ since CRs, by driving the $\rm H_3^{+}$ formation in such regions, initiate the chemical network beyond the basic $\rm H_2$ molecule (itself formed on dust grains).

The effects of  strong turbulence on a CR-controlled CO/H$_2$ ratio, and its  consequences for the CO-$''$visibility$''$ of $\rm H_2$  gas in  SF galaxies, are difficult  to  assess  within   the  static  framework  of  our  work. The necessity of a dynamical approach can be shown simply by noting that the turbulent diffusion timescales $t_{\rm diff}$ are typically  shorter than chemical-equillibrium timescales $t_{\rm ch}$, the latter approximated by the timescale needed for an H{\sc i}/H$_2$ equillibrium to be reached \citep[e.g.][]{Holl99}. Indeed for an initially atomic Cold Neutral Medium (CNM) phase (and a canonical H$_2$ formation rate on grains of $R_f=3\times 10^{-18} ({\rm T/K})^{-1/2} {\rm cm}^{3}\,{\rm s}^{-1}$) it is $t_{\rm ch}\sim 10^{7}$\,yrs, while  $t_{\rm diff}\propto 1/K_{\rm diff}$ \citep[with $K_{\rm diff}=\langle V_{\rm turb} L\rangle \sim (10^{24}-10^{25})\,{\rm cm}^2\,{\rm s}^{-1}$,][]{Xie95} ranges from $t_{\rm diff}\sim 3\times 10^{6}$\,yrs for quiescent GMCs (turbulent linewidths of $V_{\rm turb}\sim (1-5)\,{\rm km}\,{\rm s}^{-1}$) to $\sim 3\times (10^{4}-10^{5})$\,yrs for clouds typical of ULIRG environments \citep[][and references therein]{Papa04} The supersonic turbulence  that   dominates the velocity  fields of  molecular  clouds is  expected  to influence  our results  in two distinct  ways namely:  a) by  strongly ``mixing''  outer and inner cloud layers  (and thus also the respective abundances of the  various species formed in those cloud regions), and b)   by controlling  the  molecular cloud  mass fraction over the gas density range i.e.  the $\rm dM_{gas}/dn$    function \citep[][and references therein]{Vazq94,Pado02}. Early work on the first issue by \citet{Xie95} indicates that    it is the abundances established on the outer cloud layers (i.e. the CO-poor/C{\sc i}-rich ones in  CR-immersed clouds with    realistic density profiles) that will act to  ``dilute'' the inner, denser, CO-rich/C{\sc i}-poor regions. Furthermore,    the typically stronger turbulence of the outer lower-density molecular cloud regions with respect to the inner more quiescent     ones can only enhance such turbulent diffusion transfer of outer layer abundances inwards. These effects would then  further reduce the CO-visibility of turbulent molecular clouds immersed in high CR energy density backgrounds.

Regarding the second issue, we note that galaxies with high average $\rm \rho_{SFR}$ values like ULIRGs are typically also those with much higher levels of turbulence (and hence dense gas mass fractions which in turn ``fuel'' their high $\rm \rho_{SFR}$). Indeed the face-on velocity dispersion of the molecular gas disks in ULIRGs is  $\rm \sigma (V_z)$$\sim $(30-140)\,km\,s$^{-1}$ \citep{Down98}, much higher than in isolated spiral disks ($\sim $10\,km\,s$^{-1}$). As a result the underlying  $\rm dM_{gas}/dn$ would ``shift'' most of the mass of a molecular cloud towards high densities ($\geq $$10^4$\,cm$^{-3}$), and thus towards the CO-rich part of the $[n_{\rm H}-\zeta_{\rm CR}]$ parameter space shown in Fig.\ref{fig:zcr}, perhaps despite the also high average $\rm \zeta_{CR}$  expected in such environments.  Only hydrodynamical simulations of molecular clouds with such high levels of turbulence {\it and} the appropriate chemical networks for CO and H$_2$ formation and destruction can decide this issue. On the other hand it is worth pointing out that the cm-emission synchrotron disks of galaxies (rough indicators of the extend of CR propagation) are typically larger than the CO-visible molecular gas distributions. Thus starburst-originating CRs may affect the CO/H$_2$ ratio beyond the starburst  region itself, where lower-density, non-SF, molecular clouds may lie.

\subsection{Some observational tests}
\label{ssec:tests}

A mainly CR-controlled CO/H$_2$ ratio in the molecular clouds of SF galaxies has some interesting observational consequences,  which can be probed using the  JVLA and ALMA, the two interferometer arrays operating in cm and  mm/submm wavelengths with  high sensitivities. The JVLA can image  the cm synchrotron radiation emitted by  CR electrons (and thus serve  as a  probe  of CR  propagation through galactic  disks and individual molecular  clouds), while ALMA can image  the spectral lines of  various  species emanating from CR-permeated molecular  gas, to yield tests of a CR-controlled chemistry. More specifically:

\begin{enumerate}

\item Deep imaging of the cm continuum of gas-rich SF disks along with  CO  $J$=1--0, 2--1 and  C{\sc i} $J$=1--0,  2--1 line  imaging can  reveal the   distribution  of  CR-permeated molecular  gas,  and  probe  the   structural  characteristics  of  the  molecular  gas  distribution  as recovered via C{\sc i} lines versus those using the two low-$J$ CO lines.

\item Imaging of individual molecular clouds in nearby SF galaxies in cm synchrotron continuum as well as in CO $J$=1--0, 2--1 and C{\sc i} $J$=1--0, 2--1   lines in order to examine the effects of CRs in individual GMCs.

\item Any  molecular line observations constraining the average x(e) (ionization fraction)   within molecular clouds,   versus  CR  distribution (traced  by  cm  continuum imaging), can reveal whether CRs penetrate throughout GMC volumes determining their chemistry  (e.g. x(e) correlating with cm emission brightness).

\end{enumerate}

Generally in high  CR energy density environments we  expect $\rm H_2$ clouds to be marked by the usual CO-bright emission emanating from the regions that are dense enough  to retain their CO-richness, {\it yet having C{\sc i}  line  emission extending  much  further  out  than those CO-bright  regions}, tracing a CO-poor phase where CO has been destroyed by CRs. Given the elevated temperatures of the latter phase (see Figs.\ref{fig:xsec} \& \ref{fig:tgas}) we expect both the C{\sc i} $J$=1--0 ($\rm E_{10}/k_B$$\sim $24\,K) and C{\sc i} $J$=2--1 ($\rm E_{21}/k_B$$\sim $62\,K) to be substantially excited. In principle this allows C{\sc i} line imaging observations of SF disks to access a larger co-moving volume of the Universe and serve as alternative $\rm H_2$ gas distribution tracers.

\section{Conclusions}

In this paper we examined the impact of CRs on the average CO/H$_2$ ratio in $\rm H_2$ clouds.  We find that even ordinary boosts of CR energy densities, expected in numerous SF galaxies in the Universe, can destroy CO (but not $\rm H_2$), so effectively as to render potentially large  $\rm H_2$ gas reservoirs CO-poor and difficult to trace using conventional  CO $J$=1--0, 2--1, line imaging. This is a density-dependent effect, with denser regions able to retain higher  CO/H$_2$ ratios than low-density ones, for a given CR energy density (and thus CR ionization rate). This may give rise to a second order effect with $\rm H_2$ gas disks that  host vigorous SF, appearing clumpier in CO $J$=1--0, 2--1 emission than they really are, their denser (and clumpier) $\rm H_2$  gas regions having retained large enough CO/H$_2$ ratios as to remain CO-visible, unlike the CO-poor lower-density ones. Furthermore, even for CR energy density backgrounds expected in the Milky Way, low-density gas ($n_{\rm H}\sim1-3\times10^2\,{\rm cm}^{-3}$) can also be very CO-poor. The CR-induced destruction of CO leaves behind a C{\sc i}-rich  molecular gas phase, with only small amounts of C{\sc ii}. The  C{\sc i} $J$=1--0, 2--1 lines are then the only remaining effective $\rm H_2$ mass distribution tracers, encompassing both for the CO-rich and the CO-poor phase.

 Strong turbulence, often found for the molecular gas of vigorously SF galaxies, can in principle help maintain much of it CO-rich (and thus CO-line-visible), by placing most of it at high average densities. Hydrodynamical simulations of highly turbulent, CR-irradiated molecular gas, which incorporate the appropriate  CO, H$_2$ formation/destruction chemical networks, are necessary to explore this issue. On the observational front  deep C{\sc i}, low-$J$ CO line, and cm continuum imaging observations of SF galaxies and  individual molecular clouds embedded in environments of high $\rm U_{CR}$, are key in exploring a CR-regulated CO/H$_2$ ratio in $\rm H_2$ clouds. We expect the  C{\sc i} line emission  to extend well beyond that of low-$J$ CO line emission if that is indeed so.

\acknowledgments

We thank the anonymous referee for suggestions that improved a previous version of the manuscript and enlarged its scope. The work of TGB was funded by STFC grant ST/J001511/1. The work of PPP was funded by an Ernest Rutherford Fellowship. The authors thank Antonios Makrymallis for his contribution on calculating Eqn.\ref{eqn:function}. TGB and PPP thank Nikos Karafolas and Despoina Gika for their hos\,pita\,lity at Seli (and the pita in particular), Veroia, where principal ideas of the present paper have been discussed. This research has made use of NASA's Astrophysics Data System.

Parts of the calculations of the present paper used the DiRAC Data Analytic system at the University of Cambridge, operated by the University of Cambridge High Performance Computing Service on behalf of the STFC DiRAC HPC Facility (www.dirac.ac.uk). This equipment was funded by BIS National E-infrastructure capital grant (ST/K001590/1), STFC capital grants ST/H008861/1 and ST/H00887X/1, and STFC DiRAC Operations grant ST/K00333X/1. DiRAC is part of the National E-Infrastructure.

\appendix
\section{The numerical model}
\label{app}

We use {\sc 3d-pdr} \citep{Bisb12}, a three-dimensional time-dependent astrochemistry code implemented at University College London and designed for treating PDRs of arbitrary density distribution. The code has been used in various applications examining galactic and extragalactic objects \citep[i.e.][]{Offn13,Offn14,Bisb14,Gach14}. For the purposes of this work, we perform high resolution one-dimensional uniform-density clouds as discussed in \citet{Bisb12}. We use a subset of the UMIST 2012 network \citep{McEl13} consisting of 58 species and more than 600 reactions. The abundances of species relative to hydrogen we used are [He]=$8.5\times10^{-2}$, [C]=$2.69\times10^{-4}$, [O]=$4.90\times10^{-4}$, [Mg]=$3.98\times10^{-5}$, and [S]=$1.32\times10^{-5}$ which correspond to undepleted solar abundances \citep{Aspl09}. Full technical details can be found in \citet{Bisb12} and in \citet{Bisb14} for further updates.

In the simulations presented here, we use an expression for molecular line optical depths suitable for macroturbulent molecular clouds.  Thus in the expression of $\tau_{ij}$ for the line transition $ij$ of
\begin{eqnarray}
\tau_{ij}=\frac{A_{ij}c^3}{8\pi\nu_{ij}^3}\left(\frac{dV}{dr}\right)_{\rm VIR}^{-1}\left[\frac{n_jg_i-n_ig_j}{g_j}\right],
\end{eqnarray}
(where $A_{ij}$ is the Einstein coefficient, $\nu_{ij}$ is the line frequency, $n_i, $$n_j$  the populations and $g_i$, $g_j$ their corresponding statistical weights of levels $i$ and $j$), we set  $\frac{dV}{dr}$  as:
\begin{eqnarray}
\left(\frac{dV}{dr}\right)_{\rm VIR}=0.65\sqrt{\alpha}\left[\frac{n_{\rm H}}{10^3\,{\rm cm}^{-3}}\right]^{1/2}\,{\rm km}\,{\rm s}^{-1}{\rm pc}^{-1},
\end{eqnarray}
suitable for a mostly self-gravitating gas phase \citep{Papa99}, where $\alpha$ depends on the assumed density profile and $n_{\rm H}$ is the total H-nucleus density. This ensures that the optical depth of the transition $ij$ is a local quantity everywhere within the cloud, as expected for macroturbulent velocity fields. For this work we adopt   $0.65\sqrt{\alpha}=1$.

\end{document}